\documentclass[%
 reprint,
bibnotes,
 amsmath,amssymb,
 aps,
]{revtex4-1}

\usepackage{xprintlen}

\usepackage{graphicx}
\usepackage[english]{babel}
\usepackage{microtype}          
\usepackage[breaklinks=true,colorlinks=true,linkcolor=blue,urlcolor=blue,citecolor=blue]{hyperref}

\usepackage{xcolor}     
\usepackage{mathtools}
\usepackage{braket}
\usepackage{tikz}
 \usepackage{printlen}

 \usepackage{longtable}
  \usepackage{booktabs}
 \usepackage{siunitx}
 \usepackage{csquotes}

\newcommand{\op}[1]{\ensuremath{\hat{#1}}}

\usepackage{dcolumn}
\usepackage{bm}

\begin{document}
\bibliographystyle{apsrev}
\preprint{APS/123-QED}

\title{Permutation Blocking Path Integral Monte Carlo approach\\ to the Static Density Response of the Warm Dense Electron Gas}

\author{Tobias Dornheim$^{1}$}%
 \email{dornheim@theo-physik.uni-kiel.de}
\author{Simon Groth$^{1}$}
\author{Jan Vorberger$^{2}$}
\author{Michael Bonitz$^{1}$}
\affiliation{ $^1$Institut f\"ur Theoretische Physik und Astrophysik, Christian-Albrechts-Universit\"{a}t zu Kiel, D-24098 Kiel, Germany\\
$^2$Helmholtz-Zentrum Dresden-Rossendorf, D-01328 Dresden, Germany}

\date{\today}

\begin{abstract}
The static density response of the uniform electron gas is of fundamental importance for numerous applications.
Here, we employ the recently developed \textit{ab initio} permutation blocking path integral Monte Carlo (PB-PIMC) technique [T.~Dornheim \textit{et al.}, \textit{New J.~Phys.}~\textbf{17}, 073017 (2015)] to carry out extensive simulations of the harmonically perturbed electron gas at warm dense matter conditions. In particular, we investigate in detail the validity of linear response theory and demonstrate that PB-PIMC allows to obtain highly accurate results for the static density response function and, thus, the static local field correction. A comparison with dielectric approximations to our new \textit{ab initio} data reveals the need for an exact treatment of correlations. Finally, we consider a superposition of multiple perturbations and discuss the implications for the calculation of the static response function.
\end{abstract}

\maketitle

\section{Introduction}

The uniform electron gas (UEG), which is comprised of Coulomb interacting electrons in a homogeneous neutralizing background, is one of the most seminal model system in quantum many-body physics and chemistry~\cite{loos}.
In addition to the UEG's importance for, e.g., the formulation of Fermi liquid theory~\cite{quantum_theory,quantum_theory2} and the quasi-particle picture of collective excitations~\cite{pines,pines2}, accurate parametrizations of its ground state properties~\cite{vwn,perdew,pw,gori,gori2} based on \textit{ab initio} quantum Monte Carlo calculations~\cite{gs1,gs2,ortiz,ortiz2,spink} have been pivotal for the arguably unrivaled success of density functional theory simulations of real materials~\cite{ks,dft_review,dft_burke}.

The density response of the UEG to a small external perturbation as described by the density response function
is of high importance for many applications~\cite{quantum_theory}. The well-known random phase approximation (RPA)~\cite{rpa_original} provides a qualitative description for weak coupling strength (high density),
\begin{eqnarray}\label{eq:rpa}
\chi_\text{RPA}(\mathbf{q},\omega) = \frac{\chi_0(\mathbf{q},\omega)}{1-\frac{4\pi}{q^2}\chi_0(\mathbf{q},\omega)} \quad ,
\end{eqnarray}
where $\chi_0(\mathbf{q},\omega)$ denotes the density response function of the ideal (i.e., non-interacting) system.
However, since Eq.~(\ref{eq:rpa}) does not incorporate correlations beyond the mean field level, RPA breaks down even for moderate coupling.
This shortcoming is usually corrected in the form of a local field correction (LFC) $G(\mathbf{q},\omega)$~\cite{kugler1}, modifying Eq.~(\ref{eq:rpa}) to
\begin{eqnarray}
\chi_\text{LFC}(\mathbf{q},\omega) = \frac{\chi_0(\mathbf{q},\omega)}{1-\frac{4\pi}{q^2}[1-G(\mathbf{q},\omega)]\chi_0(\mathbf{q},\omega)} \quad .
\end{eqnarray}
Hence, by definition, the exact LFC contains all exchange-correlation effects beyond RPA. Common approximations for $G$ include the approaches by Singwi-Tosi-Land-Sj\"olander (STLS)~\cite{stls_original} and Vashishta and Singwi (VS)~\cite{vs_original}.
It is important to note that the accurate determination of $G(\mathbf{q},\omega)$ is an important end in itself as it can be straightforwardly utilized as input for other calculations. For example, it is directly related to the XC kernel
\begin{eqnarray} 
K_\text{xc}(\mathbf{q},\omega) = - \frac{4\pi}{q^2} G(\mathbf{q},\omega)
\end{eqnarray}
of density functional theory in the adiabatic-connection fluctuation-dissipation formulation~\cite{lu,patrick,burke2}.
This allows for the construction of a true non-local XC functional, which is a promising approach to go beyond the ubiquitous gradient approximations~\cite{dft_burke,pbe} and thereby increase the predictive capabilities of DFT. Further applications of the LFCs for current warm dense matter (WDM, see below) research include the calculation of the dynamic structure factor~\cite{fortmann1,fortmann2,collision1,collision2} as it can be obtained with X-ray Thomson scattering from a variety of systems, energy transfer rates~\cite{energy_transfer1,energy_transfer2}, the electrical and optical conductivity~\cite{cond1,cond2}, and equation of state models of ionized plasmas~\cite{equation_of1,equation_of2,equation_of3}.
Finally, we mention the construction of effective potentials both for WDM~\cite{saum1,saum2} and beyond~\cite{pseudo_potential,gravel}.

In the ground state, Moroni \textit{et al.}~\cite{moroni2} obtained accurate QMC results for the static response function [i.e., $\omega\to0$, see Eq.~(\ref{eq:static})] - and thereby the static LFC - by simulating an electron gas with a weak external harmonic perturbation~\cite{moroni1,bowen1,bowen2,senatore}.
This has allowed for a systematic assessment of the accuracy of previous approximations.
Further, the \textit{ab initio} data for the LFC have subsequently been parametrized by Corradini \textit{et al.}~\cite{cdop}, and the zero temperature limit of the static density response is well understood.

However, recently there has emerged a growing interest in matter under extreme conditions, i.e., at high density and temperature, which occurs in astrophysical objects such as brown dwarfs and planet interiors~\cite{knudson,militzer}. Furthermore, similar conditions are now routinely realized in experiments with laser excited solids~\cite{ernst} or inertial confinement fusion targets~\cite{nora,schmit,hurricane3,kritcher}. This 'warm dense matter' (WDM) regime is characterized by two parameters being of the order of unity~\cite{wdm_book}: (i) the Wigner-Seitz radius $r_s=\overline{r}/a_\text{B}$ and (ii) the reduced temperature
$\theta=k_\text{B}T/E_\text{F}$, where $\overline{r}$, $a_\text{B}$ and $E_\text{F}$ denote the mean inter particle distance, Bohr radius and Fermi energy~\cite{fermi_E}, respectively.
Naturally, accurate data for the static LFC at such extreme conditions are highly desirable. In fact, in lieu of thermodynamic data often ground state results are used at WDM conditions, which might not be appropriate~\cite{wdm_book}.

Yet, a theoretical description of warm dense electrons is notoriously hard since it must account for the nontrivial interplay of (a) the strong quantum Coulomb collisions, (b) excitation effects due to the high temperature, and (c) quantum degeneracy effects (e.g. fermionic exchange). In particular, conditions (a) and (b) rule out perturbation expansions and ground state methods, respectively, leaving thermodynamic quantum Monte Carlo methods as the most promising option.
Unfortunately, QMC simulations of degenerate electrons suffer from the fermion sign problem (FSP)~\cite{loh,troyer} so that the widespread path integral Monte Carlo (PIMC) approach~\cite{cep} is limited to small system sizes and high temperatures, preventing simulations under WDM conditions~\cite{dornheim_pop}. Despite its remarkable success in the ground state, at finite temperature, the fixed node approximation~\cite{node,bcdc} (which avoids the FSP) can lead to systematic errors exceeding $10\%$~\cite{tim3}. This unsatisfactory situation has sparked remarkable progress in the field of fermionic QMC simulations. In particular, the joint usage of two novel complementary approaches (in combination with an improved finite-size correction~\cite{dornheim_prl}) has recently allowed to obtain the first complete \textit{ab initio} description of the warm dense electron gas~\cite{dornheim_prl,groth_prl}: (i) At high density and weak to moderate coupling, the configuration PIMC (CPIMC) approach~\cite{tim_cpp15,tim1,groth}, which is formulated in Fock space and can be understood as a Monte Carlo calculation of the (exact) perturbation expansion around the ideal system, is capable to deliver exact results over a broad temperature range.
(ii) The permutation blocking PIMC (PB-PIMC) approach~\cite{dornheim,dornheim2,dornheim3} extends standard PIMC towards higher density and lower temperature and allows for accurate results in large parts of the WDM regime. In this work, we use the latter method to carry out simulations of the harmonically perturbed electron gas under warm dense matter conditions.

A brief introduction of the UEG model (Sec.~\ref{sec:UEG}) is followed by a comprehensive introduction to fermionic QMC simulations at finite temperature. In particular, we explain how the antisymmetry of the density operator leads to the fermion sign problem in standard PIMC (Sec.~\ref{sec:PIMC}), and how this is addressed by the idea of permutation blocking (Sec.~\ref{sec:pbpimc}). Further, we give a concise overview of linear response theory and how the static density response can be obtained by simulating the harmonically perturbed system (Sec.~\ref{sec:lrt}). In Sec.~\ref{sec:results}, we show extensive PB-PIMC results to investigate the dependence on the perturbation strength (\ref{sec:ptb}), the convergence with the number of imaginary time propagators (\ref{sec:prop}), and the wave vector dependence (\ref{sec:wv}), which also allows to address possible finite-size effects. Finally, in Sec.~\ref{sec:mq} we consider the response to a superposition of multiple perturbations with different wave vectors and the resulting implications for the calculation of $\chi$.

\section{Theory}
\subsection{Uniform Electron Gas\label{sec:UEG}}
The uniform electron gas is a model system of $N$ electrons in a positive homogeneous background that ensures charge neutrality. Throughout this work, we assume an unpolarized (paramagnetic) system, i.e., $N^\uparrow=N^\downarrow=N/2$ [with $\uparrow$ ($\downarrow$) denoting the number of spin-up (-down) electrons] and, thus,
\begin{eqnarray}
\xi = \frac{ N^\uparrow - N^\downarrow }{ N } = 0 \quad .
\end{eqnarray}
To alleviate the differences between a finite model system and the thermodynamic limit (finite-size effects), we employ Ewald summation for the repulsive pair interaction. Therefore, the Hamiltonian (in Hartree atomic units) is given by
\begin{eqnarray} \label{eq:H}
\hat H = -\frac{1}{2}\sum_{i=1}^{N}\nabla^2_i + \frac{1}{2}\sum_{i=1}^N\sum_{j\neq i}^N \Psi_\text{E}(\mathbf{r}_i, \mathbf{r}_j) + \frac{N}{2}\xi_\text{M} \quad ,
\end{eqnarray}
where $\Psi_\text{E}(\mathbf{r},\mathbf{s})$ and $\xi_\text{M}$ denote the Ewald pair potential and the well-known Madelung constant, see, e.g., Ref.~\cite{fraser}.

\subsection{Quantum Monte Carlo}
\subsubsection{Path Integral Monte Carlo\label{sec:PIMC}}
Throughout the entire work, we consider the canonical ensemble where the volume $V=L^3$ (with $L$ being the box length), particle number $N$ and inverse temperature $\beta=1/k_\text{B}T$ are fixed. To derive the path integral Monte Carlo formalism \cite{cep}, we consider the partition function 
\begin{eqnarray}
Z = \text{Tr}\hat \rho \quad , \label{eq:Z}
\end{eqnarray}
which is defined as the trace over the canonical density operator $\hat\rho$
\begin{eqnarray}
\hat\rho = e^{-\beta\hat H} \quad .
\end{eqnarray}
Let us temporarily restrict ourselves to distinguishable particles and re-write Eq.~(\ref{eq:Z}) in coordinate representation:
\begin{eqnarray}
Z = \int \text{d}\mathbf{R}\ \bra{\mathbf{R}} e^{-\beta\hat H} \ket{\mathbf{R}} \quad , \label{eq:R}
\end{eqnarray}
where $\mathbf{R}=\{\mathbf{r}_1,\dots,\mathbf{r}_N\}$ contains the all $3N$ particle coordinates. 
Since the matrix elements of $\hat\rho$ are not readily known, we use the group property
\begin{eqnarray}
e^{-\beta\hat H} = \prod_{\alpha=0}^{P-1} e^{-\epsilon\hat H} \quad ,
\end{eqnarray}
with $\epsilon = \beta/P$. Furthermore, we insert $P-1$ unity operators of the form $\hat 1 = \int \text{d}\mathbf{R}_\alpha \ket{\mathbf{R}_\alpha}\bra{\mathbf{R}_\alpha}$ into Eq.~(\ref{eq:R}) and obtain
\begin{eqnarray}\label{eq:path}
Z &=& \int \text{d}\mathbf{X}\ \bra{\mathbf{R}_0} e^{-\epsilon\hat H} \ket{\mathbf{R}_1}\bra{\mathbf{R}_1}\\ & & \dots \nonumber
\ket{\mathbf{R}_{P-1}}\bra{\mathbf{R}_{P-1}}e^{-\epsilon\hat H}\ket{\mathbf{R}_0} \quad ,
\end{eqnarray}
and the integration is carried out over $P$ sets of particle coordinates, $\text{d}\mathbf{X} = \text{d}\mathbf{R}_0 \dots \text{d}\mathbf{R}_{P-1}$.
We stress that Eq.~(\ref{eq:path}) is still exact. The main benefit of this re-casting is that the new expression involves $P$ density matrix elements, but at a $P$ times higher temperature. Each of these high temperature factors can now be substituted using some suitable high-$T$ approximation, e.g., the simple primitive factorization 
\begin{eqnarray}
e^{-\epsilon\hat H} \approx e^{-\epsilon\hat V}e^{-\epsilon\hat K} \quad ,
\end{eqnarray}
with $\hat V$ and $\hat K$ being the operators for the potential and kinetic contribution to the Hamiltonian, respectively, and which becomes exact in the limit $P\to\infty$~\cite{trotter}.
The resulting high-dimensional integral is then evaluated using the Metropolis algorithm~\cite{metropolis} (we employ a simulation scheme based on the worm algorithm~\cite{bon,bon2}).

\begin{figure}\vspace*{-0.56cm}
\hspace*{-0.4cm}\includegraphics[width=0.49\textwidth]{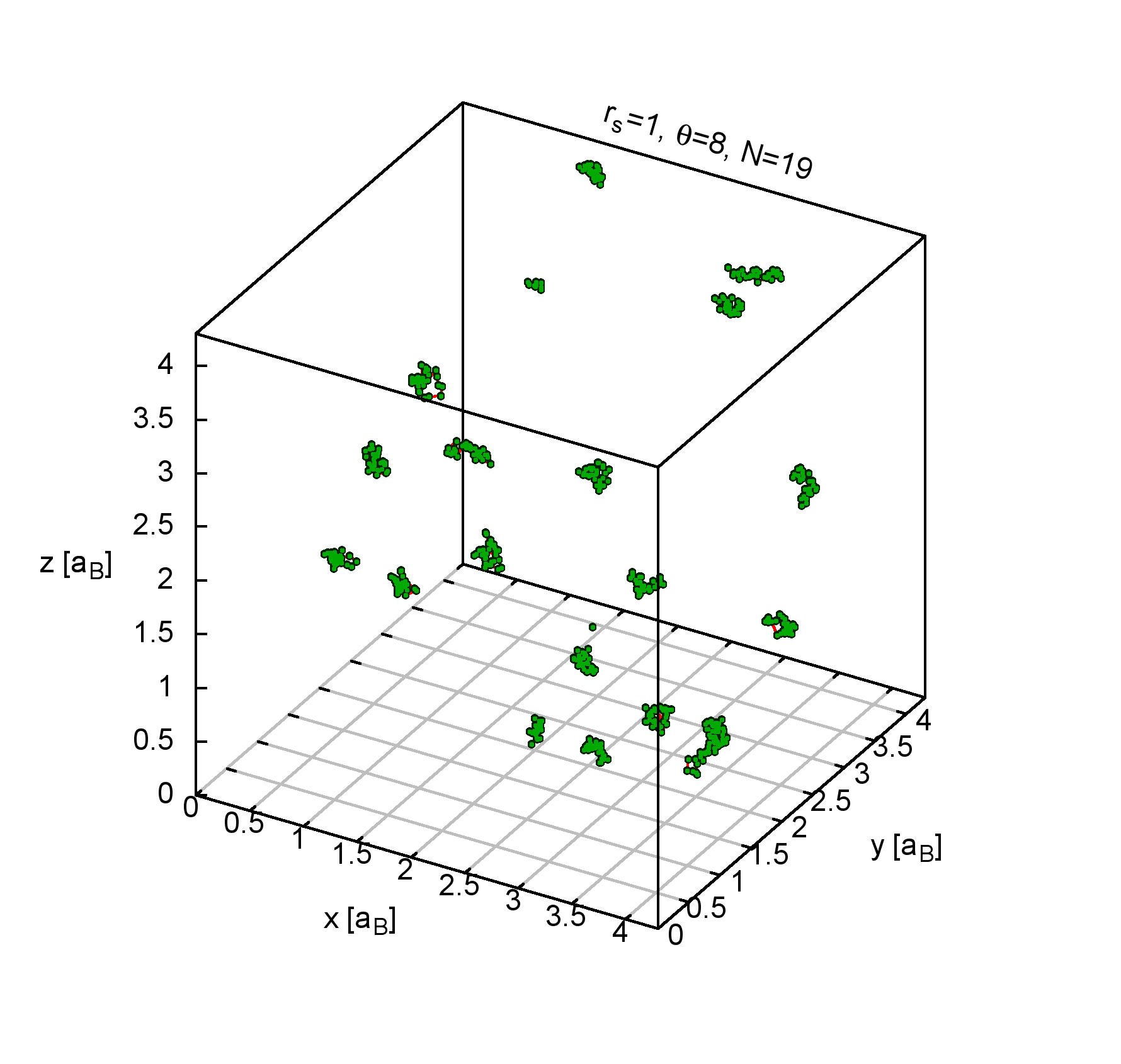}\vspace*{-0.89cm}
\hspace*{-0.4cm}\includegraphics[width=0.49\textwidth]{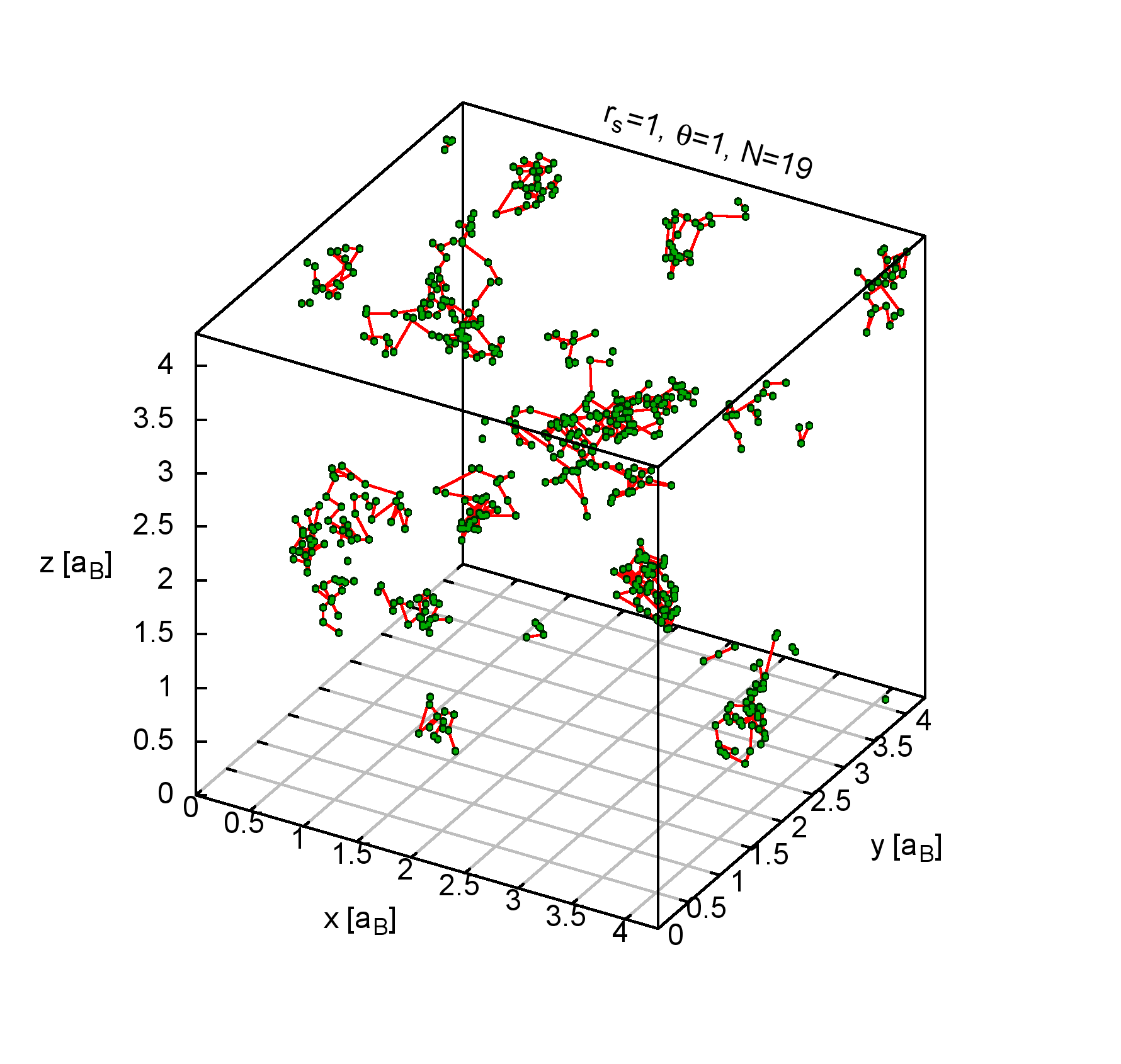}\vspace*{-0.89cm}
\hspace*{-0.4cm}\includegraphics[width=0.49\textwidth]{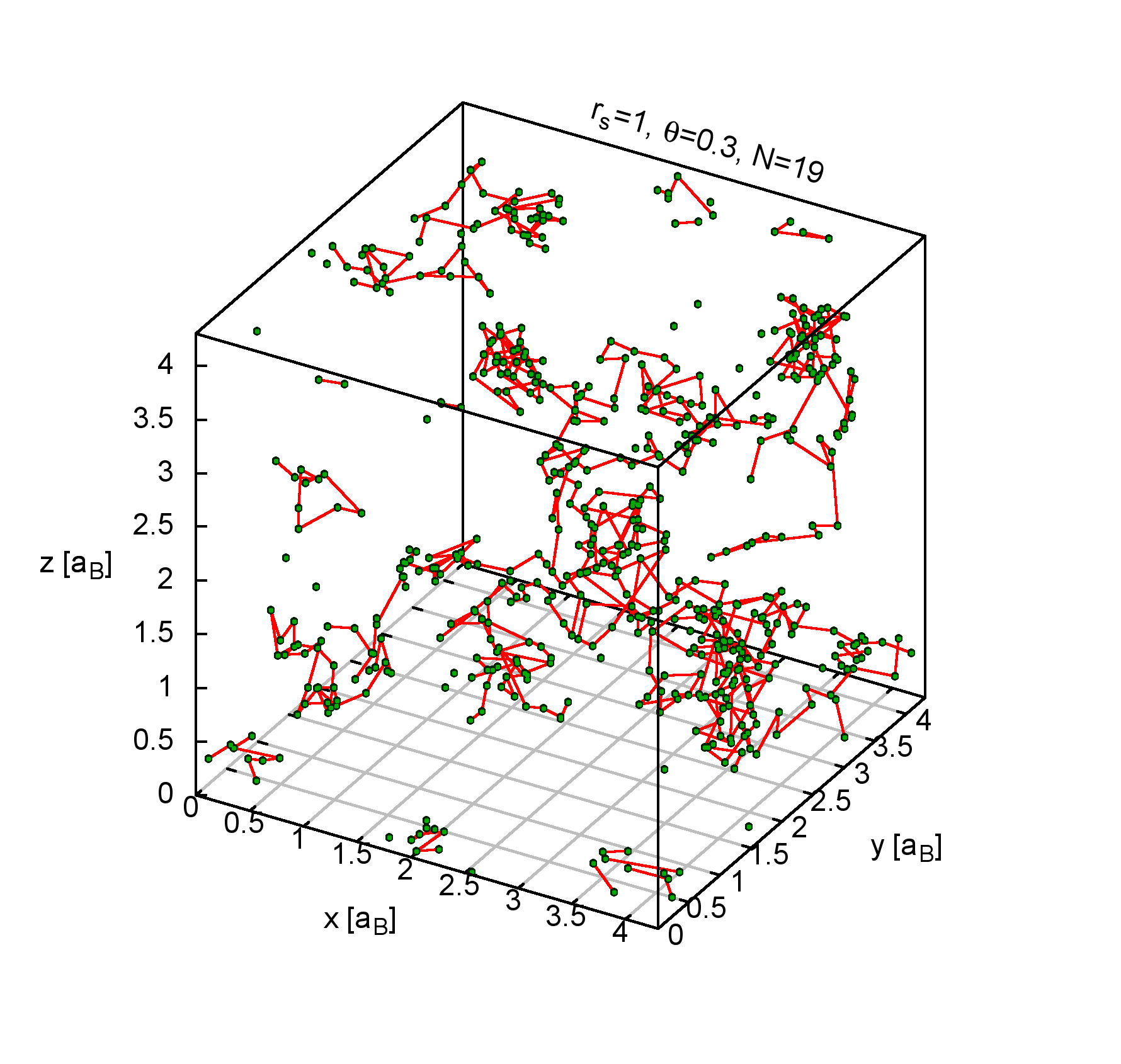}\vspace*{-0.89cm}

\caption{\label{fig:pimc_screenshots} Screen shots of standard path integral Monte Carlo simulations of the warm dense UEG for $N=19$ spin-polarized electrons, $r_s=1$, and $P=32$, with $\theta=8$ (top), $\theta=1$ (center), and $\theta=0.3$ (bottom).
}
\end{figure}

However, to simulate fermions we must extend the partition function from Eq.~(\ref{eq:R}) by the sum over all particle permutations, which, for an unpolarized system, gives
\begin{eqnarray}
Z &=& \frac{1}{N^\uparrow! N^\downarrow!}\sum_{\sigma^\uparrow\in S_{N^\uparrow}}
\sum_{\sigma^\downarrow\in S_{N^\downarrow}}
\text{sgn}(\sigma^\uparrow) \text{sgn}\left(\sigma^\downarrow\right)
\\ & & \nonumber
\int \text{d}\mathbf{R}\ \bra{\mathbf{R}}e^{-\beta\hat H}  \ket{ \hat \pi_{\sigma^\uparrow}  \hat \pi_{\sigma^\downarrow} \mathbf{R} } \quad ,
\end{eqnarray}
with $\sigma^{\uparrow,\downarrow}$ denoting particular elements from the permutation groups $S_N^{\uparrow,\downarrow}$, and $\hat \pi_{\sigma^{\uparrow,\downarrow}}$ being the corresponding permutation operators.
In practice, this leads to the occurrence of so-called exchange cycles within the PIMC simulations, which are paths incorporating more than a single particle, see Fig.~\ref{fig:pimc_screenshots}. The problem is that the sign of each configuration depends on the parity of the permutations involved which can be both positive and negative. At low temperature and high density, permutation cycles with both positive and negative signs appear with a similar frequency and, thus, the signal to noise ratio vanishes. 
This is the notorious fermion sign problem~\cite{loh,troyer}, which limits standard PIMC to weak degeneracy where fermionic exchange plays only a minor role and, therefore, precludes its application to warm dense matter~\cite{dornheim_pop}.
This is illustrated in Fig.~\ref{fig:pimc_screenshots}, where we show random configurations from standard PIMC simulations of the UEG with $N=19$ spin-polarized electrons at a density parameter $r_s=1$ and three different temperatures. Each particle is represented by $P=32$ so-called 'beads', which are connected by the (red) kinetic density matrix elements and thus form the eponymous paths.  At high temperature, $\theta=8$ (top panel), each particle is represented by a distinct, separate path and exchange cycles occur only infrequently. Therefore, the FSP is not severe and PIMC simulations are feasible. At moderate, WDM temperatures ($\theta=1$, center panel), fermionic exchange is influencing the system significantly, and multiple exchange cycles are visible in the screenshot. Since each pair exchange causes a sign change in the Monte Carlo simulation, a standard PIMC simulation is no longer feasible.
Finally, at low temperature ($\theta=0.3$, bottom panel) nearly all particles are involved in exchange cycles, and the system is dominated by the antisymmetric nature of the electrons (i.e., Pauli blocking).

\subsubsection{Permutation blocking\label{sec:pbpimc}}
The fermion sign problem is $NP$-hard~\cite{troyer} and a general solution is, at the time of this writing, not in sight. Therefore, there does not exist a single QMC method that is applicable for all parameters. Nonetheless, it is possible to go beyond standard PIMC by employing the recently introduced permutation blocking PIMC approach~\cite{dornheim,dornheim2}.
The first key ingredient is the usage of antisymmetric imaginary time propagators, i.e., determinants, which allows for a combination of positive and negative terms into a single configuration weight~\cite{det1,det2,det3}. However, while this 'permutation blocking' can indeed lead to a significant reduction of the fermion sign problem, with an increasing number of propagators $P$ this advantage quickly vanishes. For this reason, as the second key ingredient, we utilize a higher order factorization of the density matrix~\cite{ho2,ho3}
\begin{eqnarray}
 \label{cchin} e^{-\epsilon\op{H}} &\approx& e^{-v_1\epsilon\op{W}_{a_1}} e^{-t_1\epsilon\op{K}} e^{-v_2\epsilon\op{W}_{1-2a_1}} \nonumber \\ & & \times e^{-t_1\epsilon\op{K}} e^{-v_1\epsilon\op{W}_{a_1}} e^{-2t_0\epsilon\op{K}}\; ,
\end{eqnarray}
which allows for sufficient accuracy even for a small number of imaginary time slices, for the definitions of the coefficients $v_1,t_1,v_2,a_1$ and $t_0$, see Refs.~\cite{dornheim,dornheim2}.
The $\hat W$ operators correspond to modified potential terms combining the standard potential contribution $\hat V$ with double commutator terms of the form~\cite{ho3}
\begin{eqnarray}
 [[\op{V},\op{K}],\op{V}] &=& \frac{\hbar^2}{m} \sum_{i=1}^N |\mathbf{F}_i|^2\ , \\ \nonumber \mathbf{F}_i &=& -\nabla_i V(\mathbf{R})\; ,
\end{eqnarray}
where $\mathbf{F}_i$ denotes the total force on a particle '$i$'.
Finally, this allows one to obtain the PB-PIMC partition function~\cite{dornheim3}
\begin{eqnarray}
\label{finalz} Z &=&  \nonumber \frac{1}{(N_\uparrow!N_\downarrow!)^{3P}} \int \textnormal{d}\mathbf{X} \\ & &  \prod_{\alpha=0}^{P-1} \Big( e^{-\epsilon\tilde V_\alpha}e^{-\epsilon^3u_0\frac{\hbar^2}{m}\tilde F_\alpha} D_{\alpha,\uparrow}D_{\alpha,\downarrow} \Big) \; ,
\end{eqnarray}
with  $\tilde V_\alpha$ and $\tilde F_\alpha$ containing all contributions of the potential energy and the forces, respectively,
and the exchange-diffusion functions 
\begin{eqnarray}
 D_{\alpha,\uparrow} &=& \textnormal{det}(\rho_{\alpha,\uparrow} )\textnormal{det}(\rho_{\alpha A,\uparrow} )\textnormal{det}(\rho_{\alpha B,\uparrow} )\ , \\ \nonumber
  D_{\alpha,\downarrow} &=& \textnormal{det}(\rho_{\alpha,\downarrow} )\textnormal{det}(\rho_{\alpha A,\downarrow} )\textnormal{det}(\rho_{\alpha B,\downarrow} )\ .
\end{eqnarray}
Here $\rho_{\alpha,\uparrow}$ denotes the diffusion matrix of a single time slice
\begin{eqnarray}
 \rho_{\alpha,\uparrow}(i,j)  =    \lambda_{t_1\epsilon}^{-3}\sum_\mathbf{n} e^{ -\frac{\pi}{\lambda^2_{t_1\epsilon}} ( \mathbf{r}_{\alpha,\uparrow,j} - \mathbf{r}_{\alpha A,\uparrow,i}+\mathbf{n}L)^2  }\;, \label{diffusion}
\end{eqnarray}
with $\lambda_{t_1\epsilon}=\sqrt{2\pi\epsilon t_1\hbar^2/m}$ being the corresponding thermal
wavelength. Observe that Eq.~(\ref{cchin}) implies that there are three imaginary time slices for each propagator $\alpha=0,\dots,P-1$, with $\mathbf{R}_\alpha$, $\mathbf{R}_{\alpha A}$ and $\mathbf{R}_{\alpha B}$ denoting the corresponding sets of particle coordinates.

\begin{figure}\vspace*{-0.7cm}\hspace*{-0.4cm}
\includegraphics[width=0.55\textwidth]{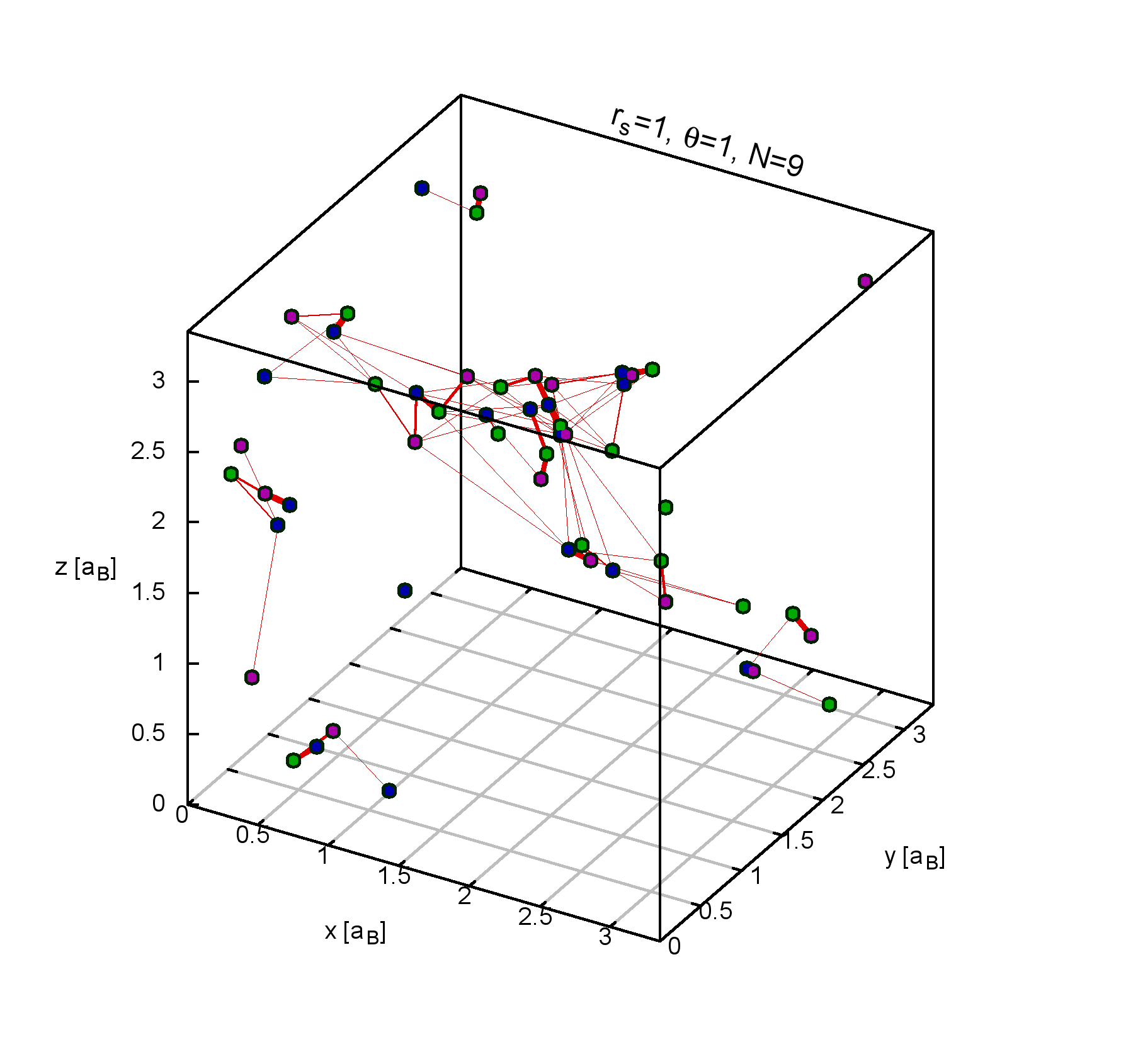}\vspace*{-0.7cm}

\caption{\label{fig:pbpimc_screenshot}Screen shot of a permutation blocking path integral Monte Carlo simulation of the UEG with $N=9$ spin-polarized electrons with $r_s=1$, $\theta=1$, and $P=2$ imaginary time propagators. The green, blue and purple points correspond to the three different kinds of time slices, see Refs.~\cite{dornheim,dornheim2,dornheim3}.
}
\end{figure}

In a nutshell, in the PB-PIMC approach, we do not have to explicitly sample each positive or negative permutation cycle. Instead, we combine configuration weights with different signs in the determinants, which results in an analytical cancellation of terms and, thus, a significantly alleviated sign problem. 
This is illustrated in Fig.~\ref{fig:pbpimc_screenshot}, where we show a random configuration from a PB-PIMC simulation of the warm dense UEG with $N=9$ spin-polarized electrons, $r_s=1$ and $\theta=1$ for $P=2$. The green, blue and purple beads correspond to the three different kinds of imaginary time slices due to the higher order factorization of the density operator, cf.~Eq.~(\ref{cchin}). In contrast to the standard PIMC configurations from Fig.~\ref{fig:pimc_screenshots}, every bead can be involved in multiple connections here. In fact, each bead is connected to all $N$ beads on the next and previous slices although the weight of the connection exponentially decreases with spatial difference, which is expressed by the different line widths of the (red) connections. Evidently, many beads of the depicted screen shot exhibit multiple visible connections, which means that a significant amount of analytical cancellation is accomplished within the determinants and, unlike standard PIMC, simulations are still feasible~\cite{dornheim_pop}.

This permutation blocking is most effective when $\lambda_{t_1\epsilon}$ is comparable (or larger) than the mean inter-particle distance. However, for $P\to\infty$ the beneficial effect vanishes and the original sign problem from standard PIMC is recovered. This plainly illustrates the paramount importance of a sophisticated higher order factorization scheme such as Eq.~(\ref{cchin}).

\subsection{Linear Response Theory\label{sec:lrt}}
In linear response theory (LRT), we consider the effect of a small external perturbation on the density of the system of interest
\begin{eqnarray}
 \hat H = \hat H_0 + \hat H_\text{ext}(t) \quad .
\end{eqnarray}
Note that, in general, $\hat H_\text{ext}(t)$ is time-dependent.
Throughout this work, the unperturbed Hamiltonian $\hat H_0$ corresponds to the UEG as introduced in Eq.~(\ref{eq:H}) and the perturbation is given by a sinusoidal external charge density of wave vector $\mathbf{q}$
\begin{eqnarray}\label{eq:ext}
 \hat H_\text{ext}(t) = 2 A\sum_{i=1}^N \text{cos}\left( \mathbf{r}_i\cdot\mathbf{q}-\Omega\ t\right)\quad,
\end{eqnarray}
which corresponds to the potential
\begin{eqnarray}
\phi_\text{ext}(\mathbf{r},t) = 2A\ \text{cos}\left(\mathbf{r}\cdot\mathbf{q}-\Omega\ t\right) \quad .
\end{eqnarray}

The standard definition of the density response function is given by
\begin{eqnarray}\label{eq:defi}
\tilde\chi(\mathbf{q},\tau) = \frac{-i}{\hbar} \braket{ \left[ \rho(\mathbf{q},\tau),\rho(-\mathbf{q},0) \right] }_0 \Theta(\tau)\ ,
\end{eqnarray}
where the expectation value is with respect to the unperturbed system. 
Note that Eq.~(\ref{eq:defi}) only depends on the time difference $\tau = t-t'$ and, due to the homogeneity of the unperturbed system, $\chi$ only depends on the modulus of the wave vector.
The corresponding Fourier transform is given by
\begin{eqnarray}\label{eq:ft}
\chi(\omega,\mathbf{q}) = \lim_{\eta\to0}\int_{-\infty}^\infty \text{d}\tau\ 
e^{(i\omega-\eta)\tau}\tilde\chi(\mathbf{q},\tau)\ .
\end{eqnarray}

Throughout this work, we restrict ourselves to the static limit that is defined as
\begin{eqnarray}\label{eq:static}
\lim_{\omega\to0}\chi(\omega,\mathbf{q}) = \chi(\mathbf{q}) \quad ,
\end{eqnarray}
i.e., the response of the electron gas to a time-independent external perturbation
\begin{eqnarray}
\phi_\text{ext}(\mathbf{r}) = 2A\ \textnormal{cos}(\mathbf{r}\cdot\mathbf{q})\ ,
\end{eqnarray}
and, henceforth, the $\omega$-dependence is simply dropped.
More precisely, the physical interpretation of $\chi(\mathbf{q})$ is the description of the density response (i.e., the induced charge density $\rho_\text{ind}(\mathbf{q})$) due to the external charge density $\rho_\text{ext}(\mathbf{q})$
\begin{eqnarray}\label{eq:chi}
\rho_\text{ind}(\mathbf{q}) = \rho_\text{ext}(\mathbf{q})\frac{4\pi}{q^2}\chi(\mathbf{q})\ .
\end{eqnarray}
The external density follows from the Poisson equation as
\begin{eqnarray}
 \rho_\textnormal{ext}(\mathbf{r}) &=& - \frac{1}{4\pi} \nabla^2 \phi_\textnormal{ext}(\mathbf{r})\\ &=& \nonumber \frac{q^2}{4\pi} \phi_\textnormal{ext}(\mathbf{r}) = \frac{q^2}{4\pi} 2A\ \textnormal{cos}(\mathbf{r}\cdot \mathbf{q}) \\
 \Rightarrow \rho_\textnormal{ext}(\mathbf{q}) &=& \frac{q^2}{2\pi} \frac{A}{(2\pi)^3}\int \textnormal{d}\mathbf{r}\ \nonumber e^{-i\mathbf{k}\cdot\mathbf{r}} \left( \frac{ e^{i\mathbf{q}\cdot\mathbf{r}} + e^{-i\mathbf{q}\cdot\mathbf{r}} }{2} \right) \\
 &=& \frac{q^2 A}{4\pi}\left( \delta_{\mathbf{k},\mathbf{q}} + \delta_{\mathbf{k},\mathbf{-q}} \right)\ ,
\end{eqnarray}
and the induced density is the difference between the perturbed and unperturbed systems:
\begin{eqnarray}\label{eq:ex}
 \rho_\text{ind}(\mathbf{q}) &=& \braket{\hat\rho_\mathbf{q}}_A - \braket{\hat\rho_\mathbf{q}}_0 \\
&=& \nonumber \frac{1}{V}\left<\sum_{j=1}^Ne^{-i\mathbf{q}\cdot\mathbf{r}_j} \right>_A\ ,
\end{eqnarray}
where we made use of the fact that $\braket{\hat\rho_\mathbf{q}}_0=0$.
Thus, it holds
\begin{eqnarray}\label{eq:weg1}
\chi(\mathbf{q}) = \frac{1}{A} \braket{\hat\rho_\mathbf{q}}_A \quad .
\end{eqnarray}

In order to obtain the desired static density response function, we carry out multiple QMC simulations for each wave vector $\mathbf{q}=2\pi L^{-1} (a,b,c)^T$ (with $a,b,c\in\mathbb{Z}$) for different values of $A$
and compute the expectation value from Eq.~(\ref{eq:ex}).
For sufficiently small $A$, $\braket{\hat\rho_\mathbf{q}}_A$ is linear with respect to $A$ with $\chi(\mathbf{q})$ being the slope.

Another way to obtain the response function from the QMC simulation of the perturbed system is via the perturbed density profile in coordinate space:
\begin{eqnarray}\label{eq:weg2}
\braket{n(\mathbf{r})}_A = n_0 + 2A\ \text{cos}\left(\mathbf{q}\cdot\mathbf{r}\right)\chi(\mathbf{q})\ .
\end{eqnarray}
In practice, we compute the lhs.~of Eq.~(\ref{eq:weg2}) using QMC and perform a fit of the rhs.~with $\chi(\mathbf{q})$ being the only free parameter. Naturally, in the linear response regime both ways to obtain $\chi(\mathbf{q})$ are equal.

For completeness, we mention that the dynamic response can be obtained in a similar fashion by considering explicitly time dependent perturbations, e.g., using non-equilibrium Green function techniques~\cite{negf1,negf2} for quantum systems or molecular dynamics \cite{md1,md2} in the classical case.

A second strategy to compute the density response from thermodynamic QMC simulations in LRT is by considering imaginary-time correlation functions (ITCF) of the unperturbed system. In particular, the static response function can be obtained from the fluctuation dissipation theorem~\cite{bowen1}
\begin{eqnarray}\label{eq:acf}
\chi(\mathbf{q}) = - \frac{1}{V} \int_0^\beta \textnormal{d}\tau\ \braket{\rho(\mathbf{q},\tau)\rho(\mathbf{-q},0)}_0\ ,
\end{eqnarray}
as an integral over the imaginary time $\tau$. If one is solely interested in the linear response of the system, invoking Eq.~(\ref{eq:acf}) constitutes the superior strategy since all $\mathbf{q}$-vectors can be computed from a single simulation. However, this requires a QMC estimation of the ITCF on a sufficient $\tau$-grid, which is straightforward in standard PIMC where $P>100$ is not an obstacle. For PB-PIMC, simulations are only possible for a small number of imaginary-time propagators (typically $P\lesssim4$), see Sec.~\ref{sec:pbpimc}, which precludes the evaluation of Eq.~(\ref{eq:acf}). Nevertheless, we stress that it is only the permutation blocking idea that allows to carry out simulations at warm dense matter conditions in the first place, since standard PIMC simulations are not feasible due to the FSP. In addition, the application of an external perturbation allows to go beyond LRT and to consider arbitrarily strong perturbation strengths.

\section{Results\label{sec:results}}

\subsection{Dependence on Perturbation strength\label{sec:ptb}}
\begin{figure}
\includegraphics[width=0.39\textwidth]{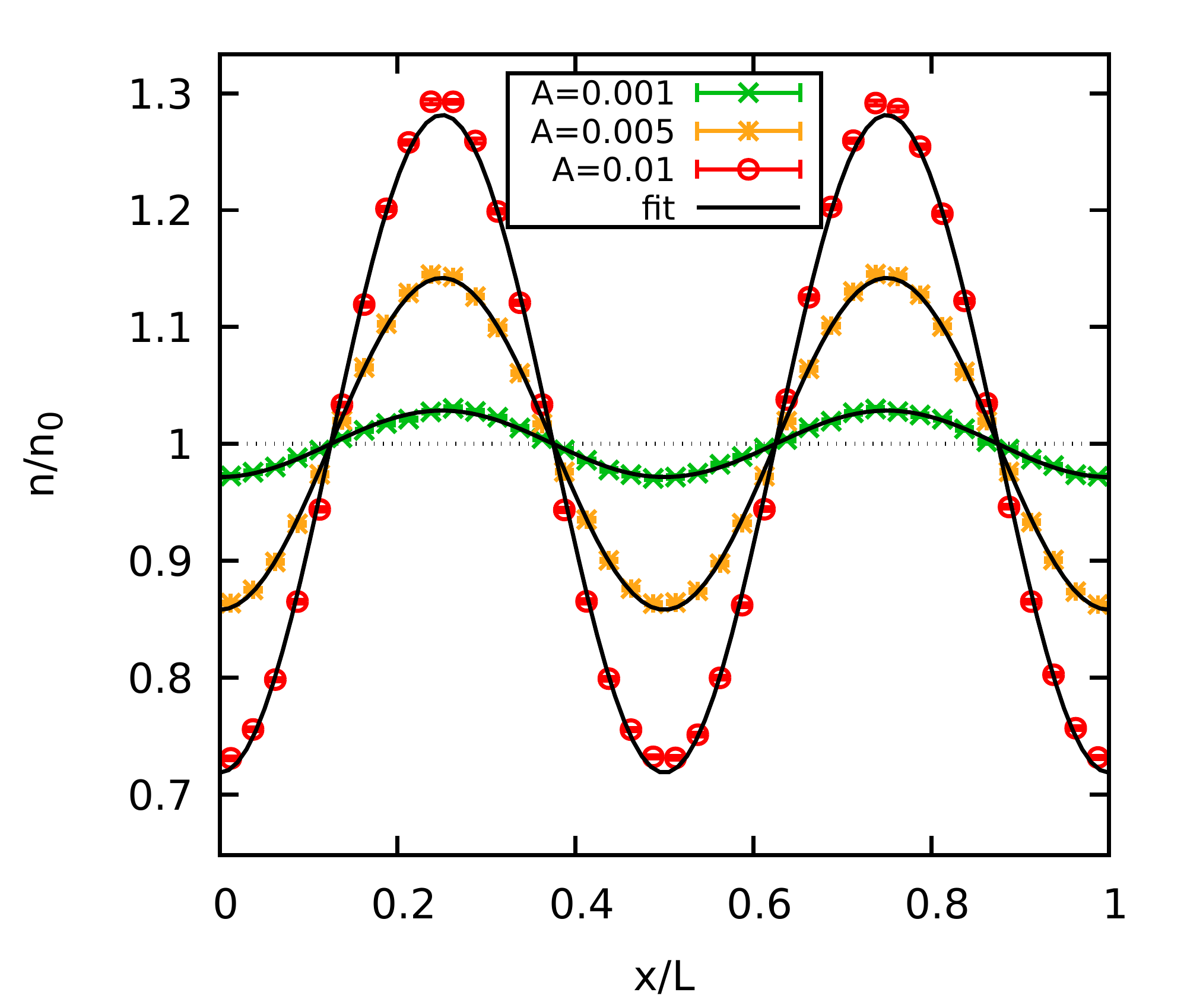}
\includegraphics[width=0.39\textwidth]{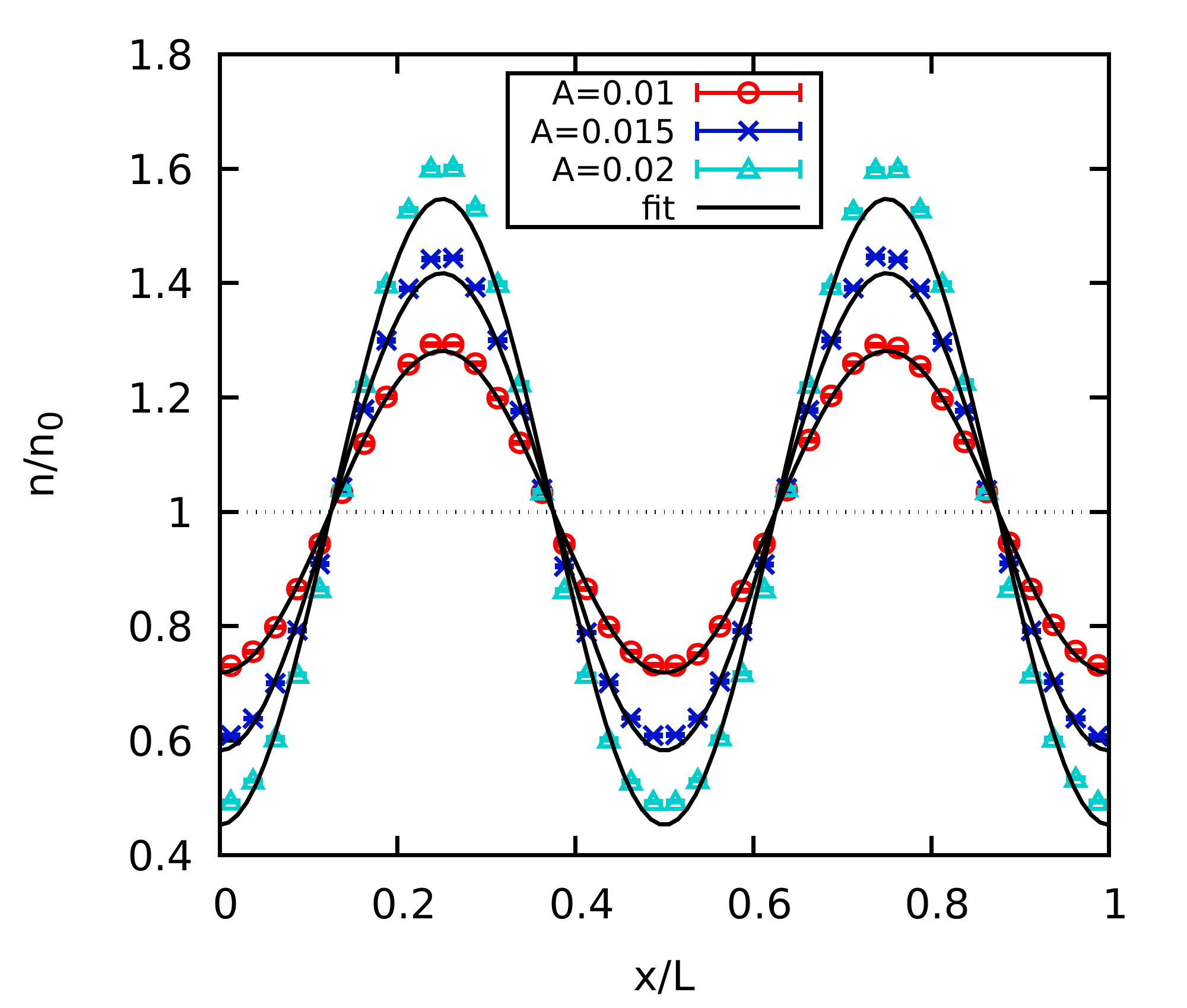}
\includegraphics[width=0.39\textwidth]{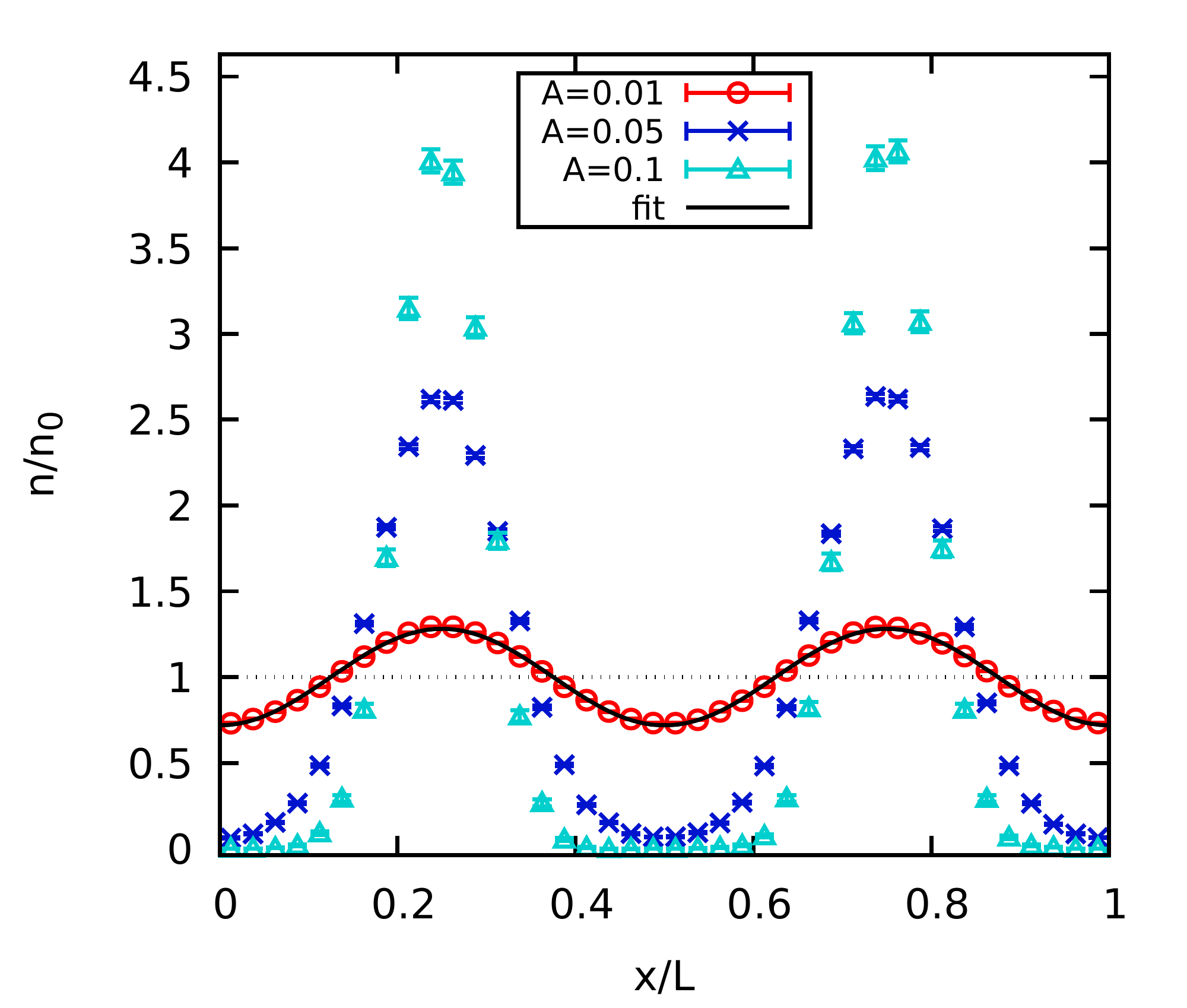}

\caption{\label{fig:1_density_qx2}Density profiles along the $x$-direction for $N=54$, $r_s=10$, and $\theta=1$. Shown are PB-PIMC results for $P=4$ with $\mathbf{q}=2\pi L^{-1}(2,0,0)^T$ and weak (top), medium (center) and strong (bottom) perturbations. The black lines correspond to fits according to Eq.~(\ref{eq:weg2}).
}
\end{figure}

Let us start our investigation of the harmonically perturbed electron gas by considering the dependence on the perturbation amplitude $A$. In Fig.~ \ref{fig:1_density_qx2}, we show PB-PIMC results for the density profile along the $x$-direction for $N=54$ unpolarized electrons at $r_s=10$ and $\theta=1$ for the perturbation wave vector $\mathbf{q}=2\pi L^{-1}(2,0,0)^T$. In the top panel, the depicted $A$-values are relatively small. The  black lines correspond to fits according to Eq.~(\ref{eq:weg2}). Evidently, for $A=0.001$ and $A=0.005$ those curves are in perfect agreement with the QMC results, which indicates that here the linear response theory is accurate. In contrast, for $A=0.01$ significant (although small, $\Delta A/A\sim1\%$) deviations appear, which are most pronounced around the minima and maxima. 
In the center panel, we systematically increase $A$ up to a factor two. Clearly, with increasing perturbation amplitude the deviations between the exact QMC results and the cosine-fit predicted by LRT become more severe, as it is expected.
Finally, in the bottom panel we show the density profiles for even larger perturbations. Eventually, the external potential becomes the dominating feature, resulting in a strongly inhomogeneous electron gas. For the largest depicted perturbation, $A=0.1$, there appear two distinct shells with a vanishing density in between.

\begin{figure}
\includegraphics[width=0.41\textwidth]{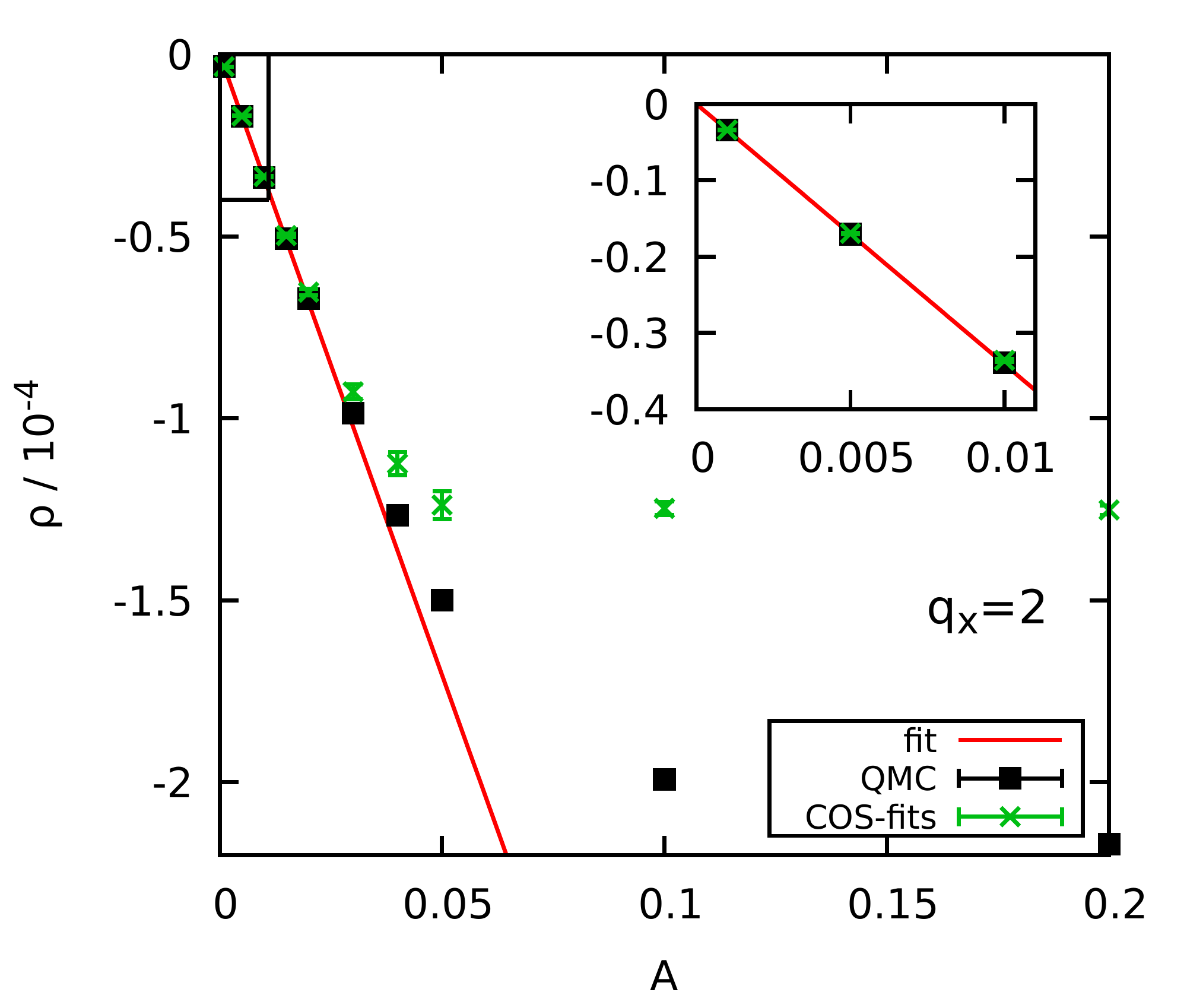}
\includegraphics[width=0.41\textwidth]{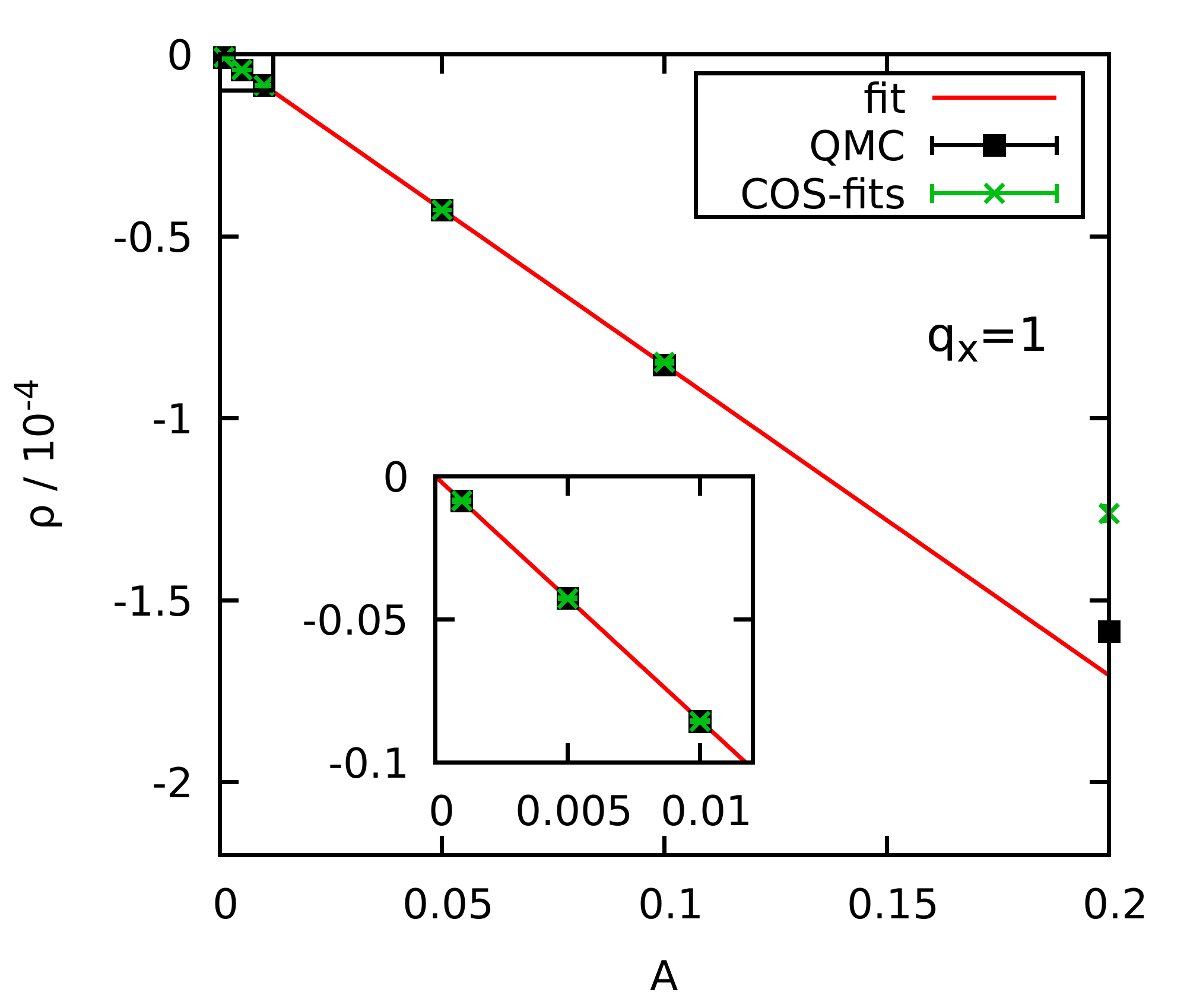}

\caption{\label{fig:1_fitplot_qx2}Induced density modulation for $N=54$, $r_s=10$, and $\theta=1$. Shown are PB-PIMC results for $P=4$ with $\mathbf{q}=2\pi L^{-1}(q_x,0,0)^T$ [$q_x=2$ (top) and $q_x=1$ (bottom)] directly computed from QMC, cf.~Eq.~(\ref{eq:ex}), and from fits according to Eq.~(\ref{eq:weg2}).
}
\end{figure}

To systematically investigate the effect of the perturbation amplitude on our QMC estimation of the static response function $\chi(\mathbf{q})$, we show results in Fig.~\ref{fig:1_fitplot_qx2} for the induced density $\rho_\text{ind}(\mathbf{q})$ for the same system and two different wave vectors, $\mathbf{q}=2\pi L^{-1}(q_x,0,0)^T$ with $q_x=2$ (top panel) and $q_x=1$ (bottom panel).
The black squares correspond to the direct QMC results, cf.~Eq.~(\ref{eq:ex}), and the green crosses have been obtained by performing a cosine-fit to the density profiles according to Eq.~(\ref{eq:weg2}). The red lines depict a linear fit to the black squares for $A<0.01$.
First and foremost, we observe a perfect agreement between the direct QMC results and the cosine-fits for small $A$ as predicted by the linear response theory. Even for $A=0.01$, where the cosine-fit exhibits significant deviations to the density profile from QMC, we find perfect agreement between the black and green points and also to the fit.
With increasing $A$, however, the assumptions of linear response theory are no longer valid. Interestingly, the $\rho$-values obtained from the cosine-fit exhibit significantly larger deviations to the linear response prediction (red line) than the direct QMC results. For example, at $A=0.05$ the deviation of the green points is twice as large as for the black squares.

In the bottom panel of Fig.~\ref{fig:1_fitplot_qx2}, the same information is shown for a smaller wave vector, $q_x=1$. Firstly, we observe a significantly smaller density response [cf.~Fig.~\ref{fig:kachel_mann}]. This, in turn, means that linear response theory is accurate up to much larger $A$-values as the system only weakly reacts to such an external perturbation.

\begin{figure}
\includegraphics[width=0.41\textwidth]{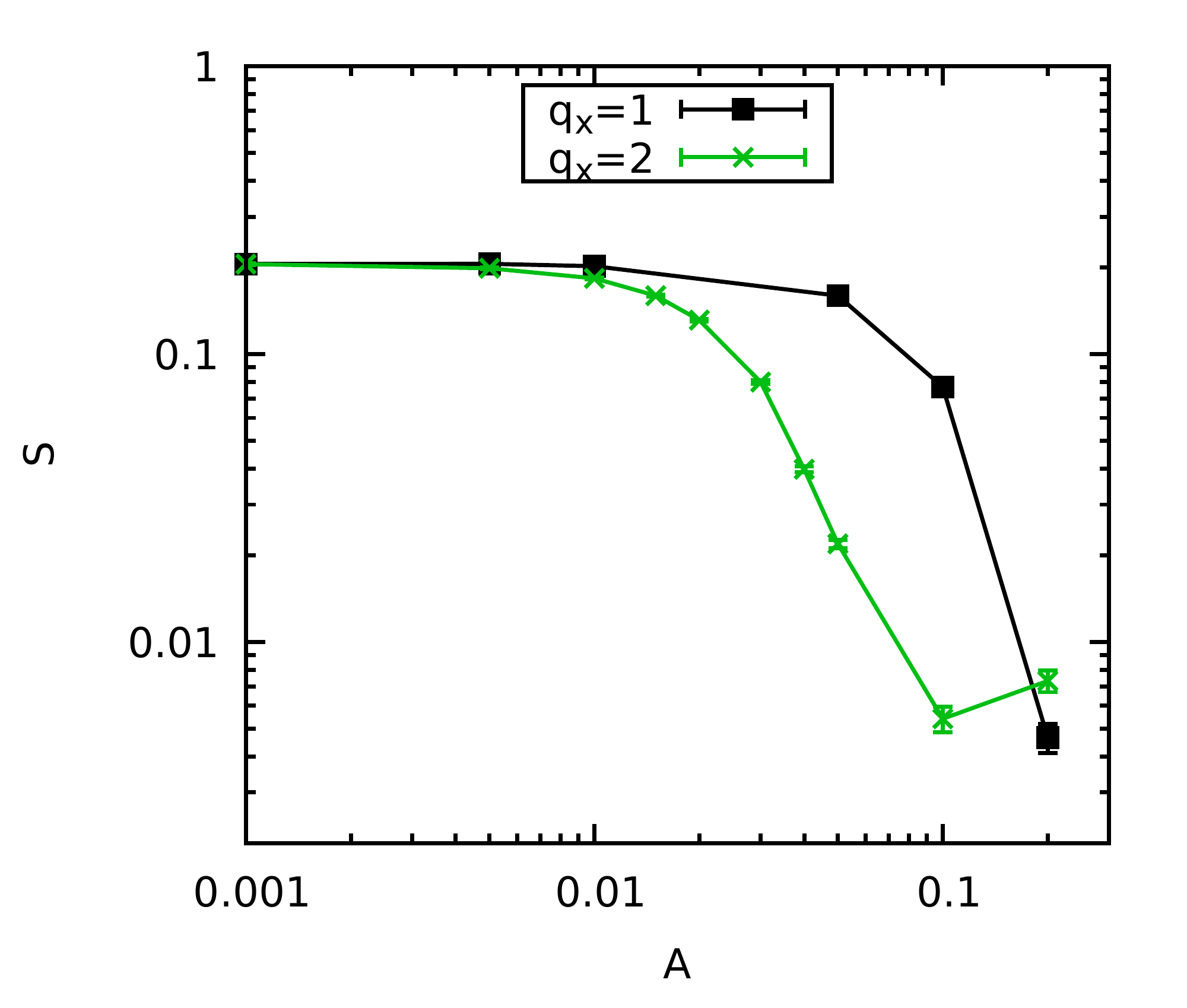}
\includegraphics[width=0.41\textwidth]{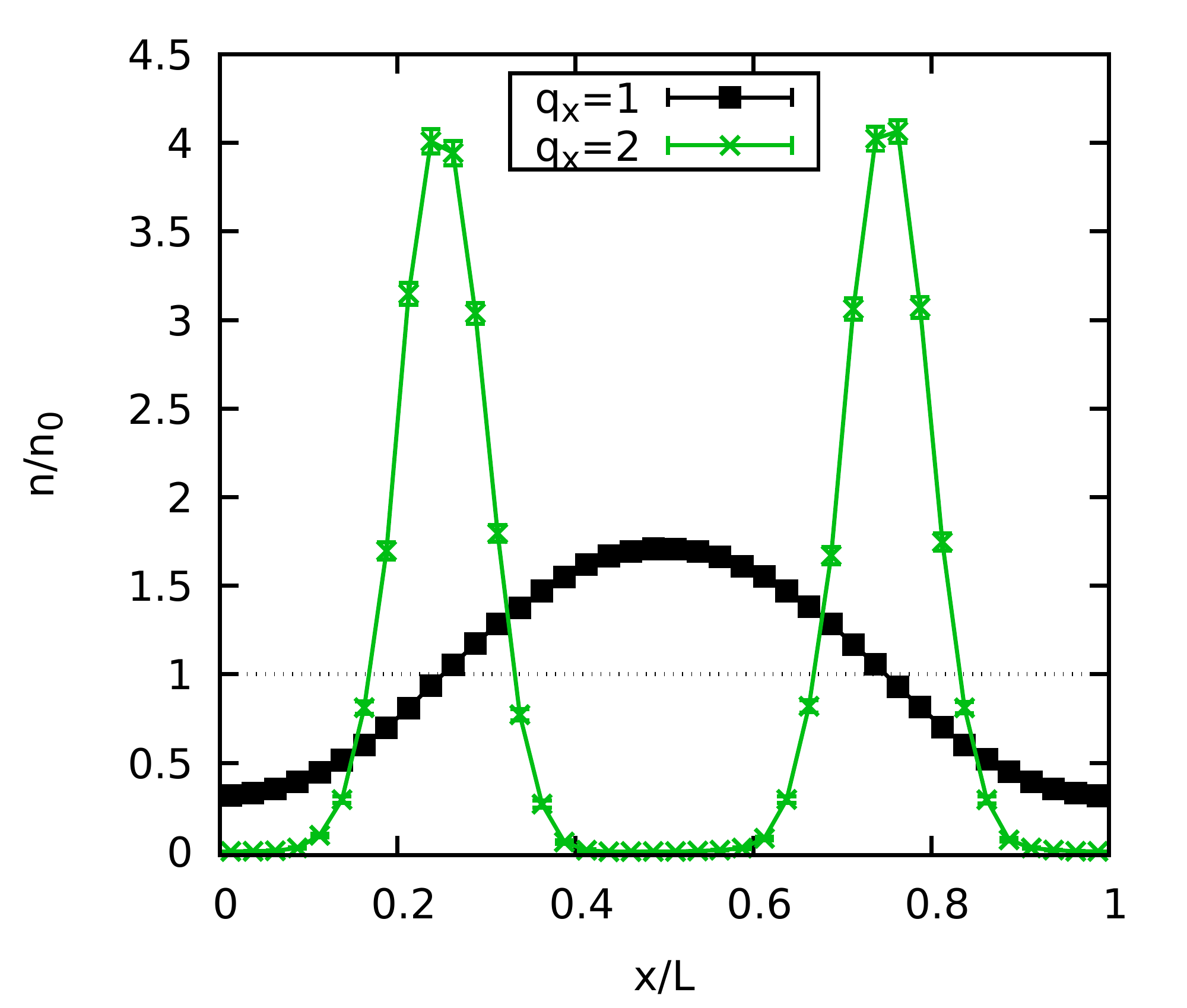}

\caption{\label{fig:1_sign_qx2} Average sign for $N=54$, $r_s=10$, and $\theta=1$ (top). Shown are PB-PIMC results for $P=4$ with $\mathbf{q}=2\pi L^{-1}(q_x,0,0)^T$. 
Corresponding density profiles along $x$-direction for $A=0.1$ (bottom).
}
\end{figure}

To further illustrate this point, in the top panel of Fig.~\ref{fig:1_sign_qx2} we show the corresponding average signs from the QMC simulations for both wave vectors investigated in Fig.~\ref{fig:1_fitplot_qx2}. 
For small perturbations, $S$ is equal for both $\mathbf{q}$ and approaches the result for the unperturbed system. With increasing $A$, the system becomes more inhomogeneous, i.e., there appear regions of increased (and also decreased) density, see the bottom panel of Fig.~\ref{fig:1_sign_qx2} where we show the corresponding density profiles for strong perturbations, $A=0.1$. This, in turn, leads to increased fermionic exchange, resulting in a significantly decreased average sign in our PB-PIMC simulations. Since the density response is more pronounced for $q_x=2$, here $S$ exhibits a faster decrease in dependence of $A$.
We conclude that PB-PIMC (and also standard PIMC) simulations of the inhomogeneous electron gas are significantly more computationally demanding than simulations of the UEG at equal conditions. Nevertheless, this is of no consequence for the determination of the static response function as this is only possible for $A$-values that are sufficiently small for the linear response theory to deliver an accurate description, i.e., systems that are close to the uniform case.

\subsection{Convergence with propagators\label{sec:prop}}
As discussed in Sec.~\ref{sec:pbpimc}, PB-PIMC crucially relies on the higher order factorization of the density operator, Eq.~(\ref{cchin}), to allow for sufficient accuracy with only few imaginary time propagators.
In the following section, this situation is investigated in detail.

\begin{figure}
\includegraphics[width=0.42\textwidth]{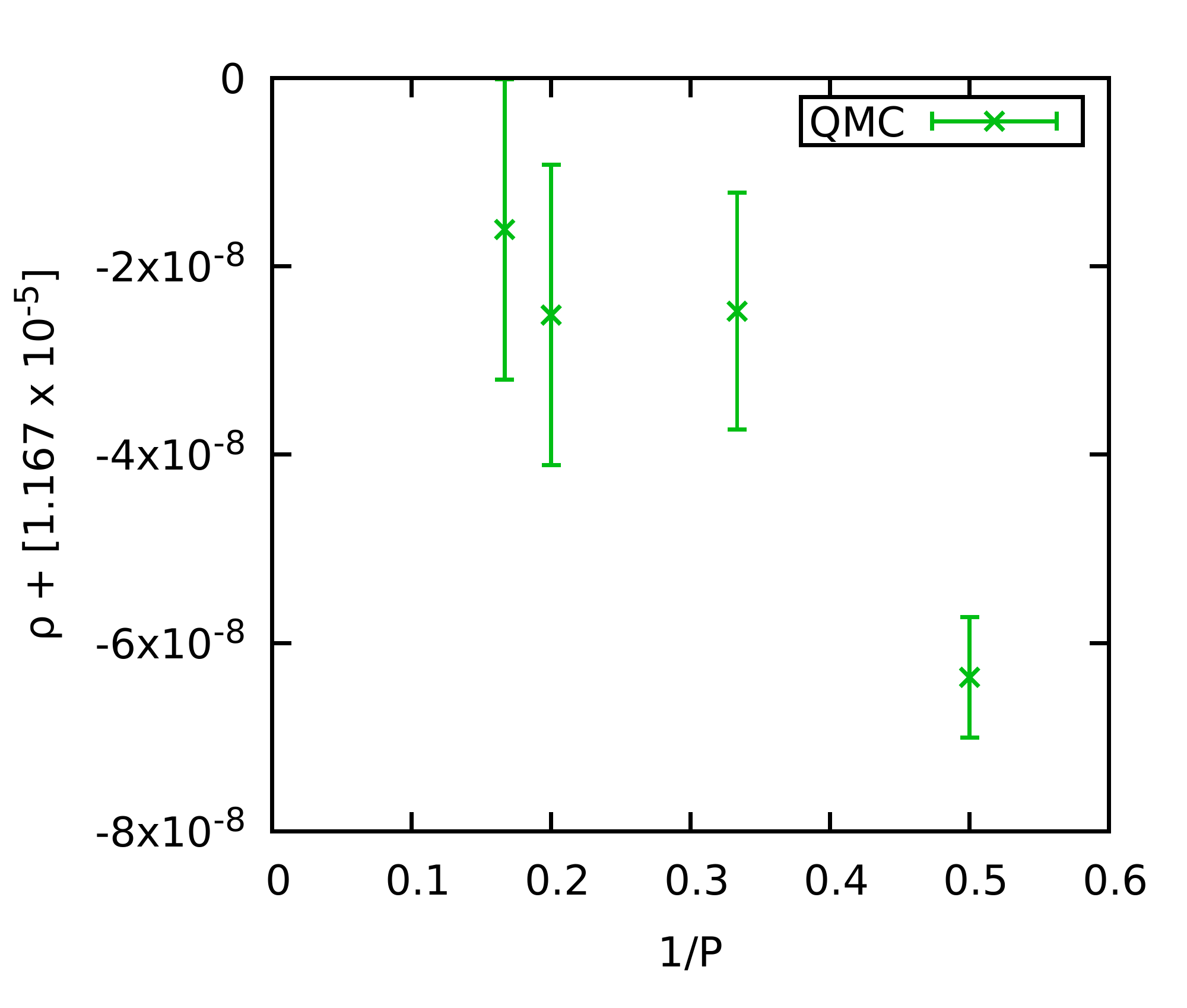}
\includegraphics[width=0.42\textwidth]{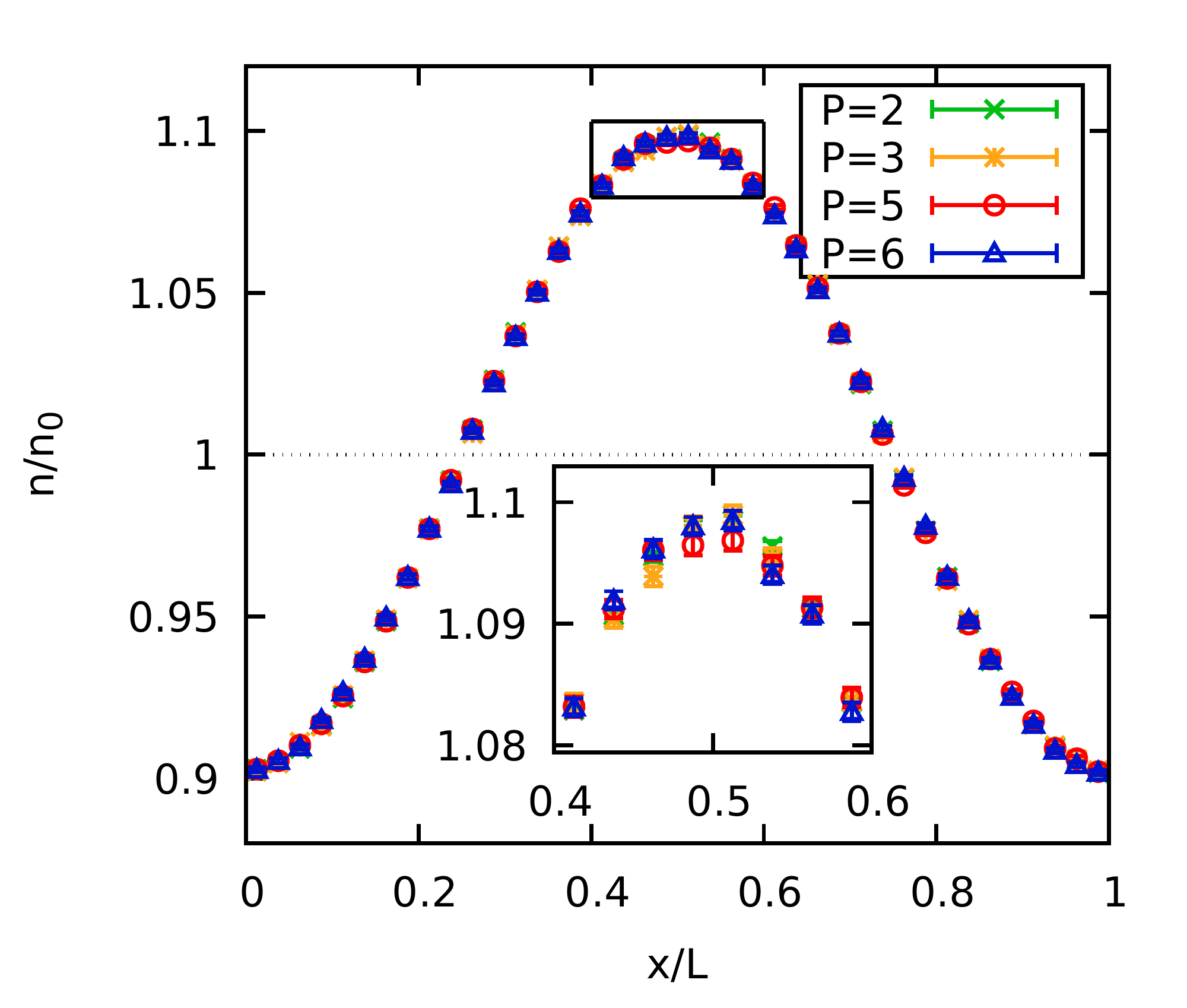}


\caption{\label{fig:propagator_small}Convergence with number of propagators $P$ for $N=34$, $r_s=10$, and $\theta=1$ with a perturbation of wave vector $\mathbf{q}=2\pi L^{-1} (1,0,0)^T$
and amplitude $A=0.01$. Shown are QMC results for the density matrix (top) and the density profile along $x$-direction (bottom).
}
\end{figure}

In the top panel of Fig.~\ref{fig:propagator_small}, we plot direct QMC results for the induced density for the unpolarized UEG with $r_s=10$, $\theta=1$, and $N=34$ electrons versus the inverse number of propagators $P^{-1}$. The perturbation is given by the wave vector $\mathbf{q}=2\pi L^{-1}(1,0,0)^T$ and amplitude $A=0.01$, which is well within the linear response regime. Evidently, only the result for $\rho$ with $P=2$ propagators significantly deviates from the rest and, for the $P=4$ propagators used above, the PB-PIMC results are converged within the statistical uncertainty.
The bottom panel shows the corresponding density profiles along the $x$-direction. Here, even the results for only $P=2$ propagators exhibits no significant deviations to the other curves.

\begin{figure}
\includegraphics[width=0.42\textwidth]{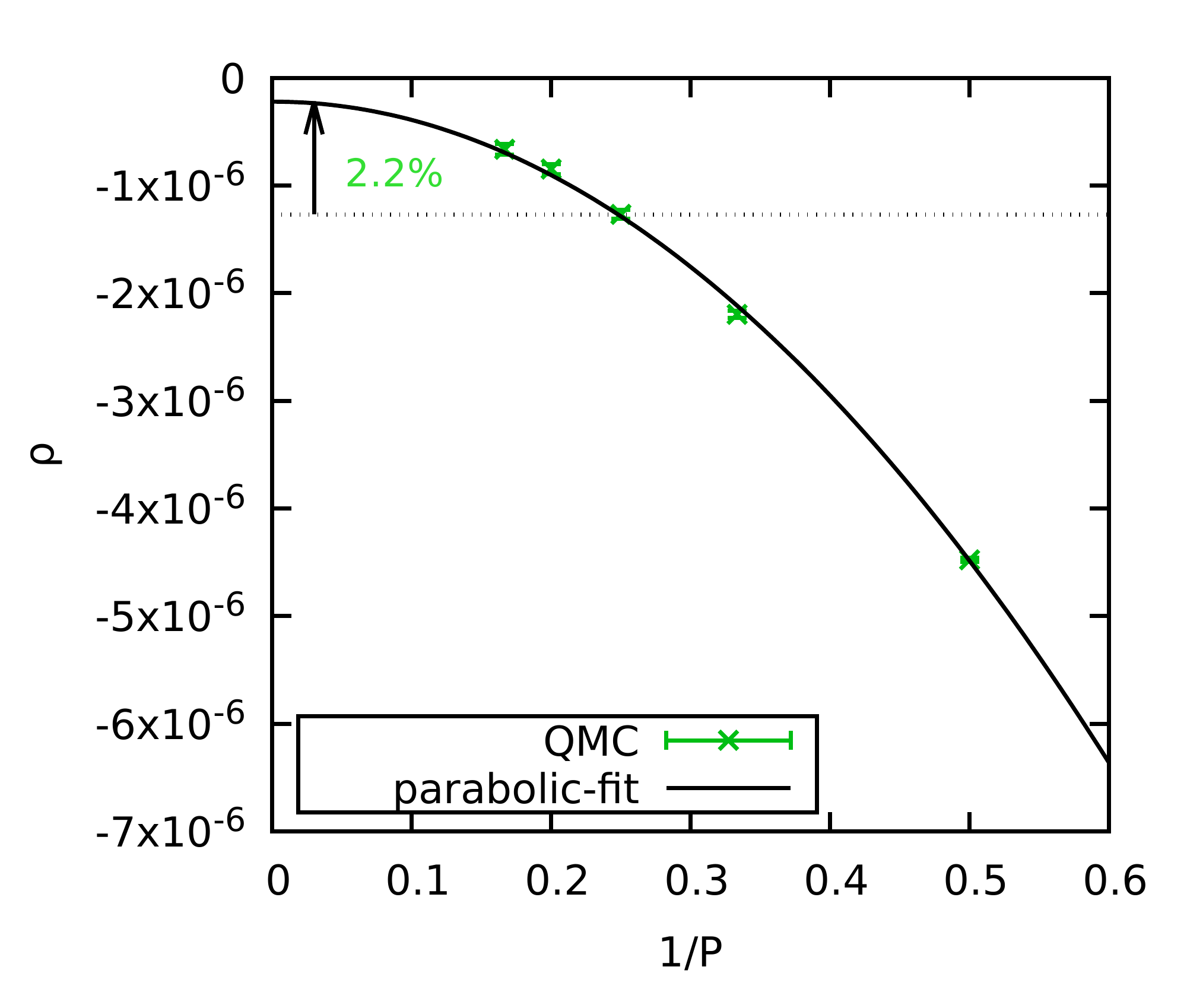}
\includegraphics[width=0.42\textwidth]{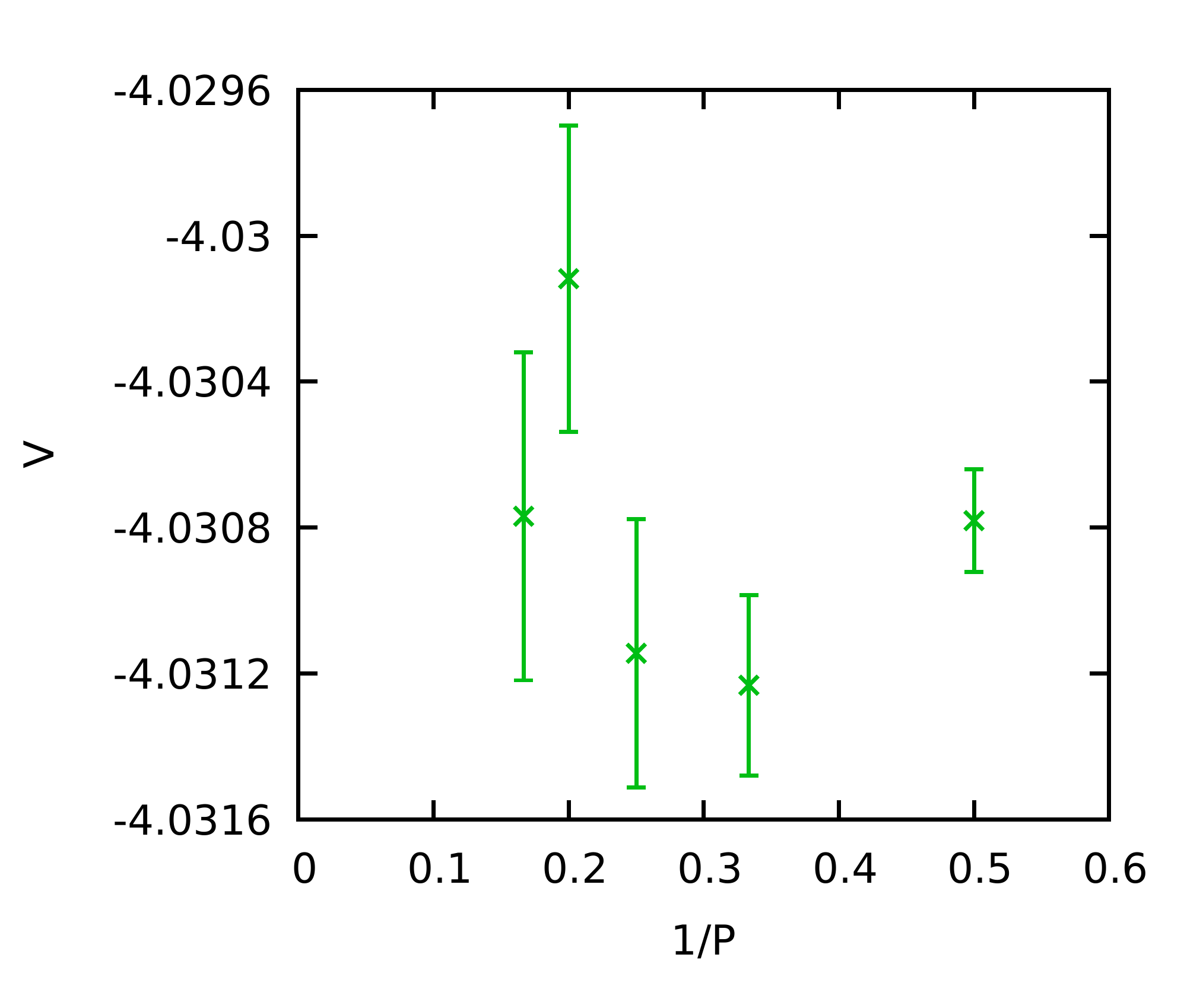}


\caption{\label{fig:propagator_large}Convergence with number of propagators $P$ for $N=54$, $r_s=10$, and $\theta=1$ with the perturbation of wave vector $\mathbf{q}=2\pi L^{-1} (5,0,0)^T$
and amplitude $A=0.01$. Shown are QMC results for the density matrix (top) and the potential energy, i.e., the sum of Ewald interaction and external field (bottom).
}
\end{figure}

As a second example, in Fig.~\ref{fig:propagator_large} we consider the same system as in Fig.~\ref{fig:propagator_small}, but with $N=54$ electrons and a larger wave vector for the perturbation, $\mathbf{q} = 2\pi L^{-1}(5,0,0)^T$. In the top panel, we again show direct QMC results for $\rho$ in dependence of the inverse number of propagators. However, in contrast to the data depicted in Fig.~\ref{fig:propagator_small}, here we see significant differences for different $P$. The black line corresponds to a parabolic fit of the form
\begin{eqnarray}\label{eq:parabolic}
\rho(P^{-1}) = a + \frac{b}{P^2} \quad ,
\end{eqnarray}
which reproduces all QMC results within error bars. Nevertheless, we stress that the functional form in Eq.~(\ref{eq:parabolic}) has been empirically chosen and does merely serve as a guide to the eye since, for large $P$, the propagator error is expected to exhibit a fourth-order decay, see Ref.~\cite{ho3} for a comprehensive discussion. 
Evidently, for $P=4$ there occurs a systematic bias of $\Delta\rho/\rho\approx2\%$ at such a large wave vector.
This is reflected in the increasing error bars towards large $\mathbf{q}$ in the wave vector dependence plot, i.e., Fig.~\ref{fig:kachel_mann}, and can be understood as follows:
The propagator error is a direct consequence of the non-commuting of the kinetic ($\hat K$) and potential ($\hat V$) contributions of the Hamiltonian. The larger the wave vector $\mathbf{q}$, the faster the spatial variations of the external potential and, because $\hat K\propto\nabla^2$, the larger the error terms, which involve nested commutators of $\hat K$ and $\hat V$.

The bottom panel of Fig.~\ref{fig:propagator_large} shows the corresponding results for the total potential energy, i.e., the sum of the Ewald interaction and the external perturbation. Evidently, no deviations can be resolved within the given statistical uncertainty, even for $P=2$ propagators. This is similar to previous findings for the unperturbed UEG~\cite{dornheim2,dornheim3} and reflects the circumstance that for $V$ the particle interaction dominates. In stark contrast, the induced density $\rho$ is particularly sensitive to the small external perturbation which, as explained above, requires a larger number of propagators to be sufficiently incorporated.

\subsection{Wave vector dependence of $\chi(\mathbf{q})$ and finite size effects\label{sec:wv}}

\begin{figure}
\includegraphics[width=0.42\textwidth]{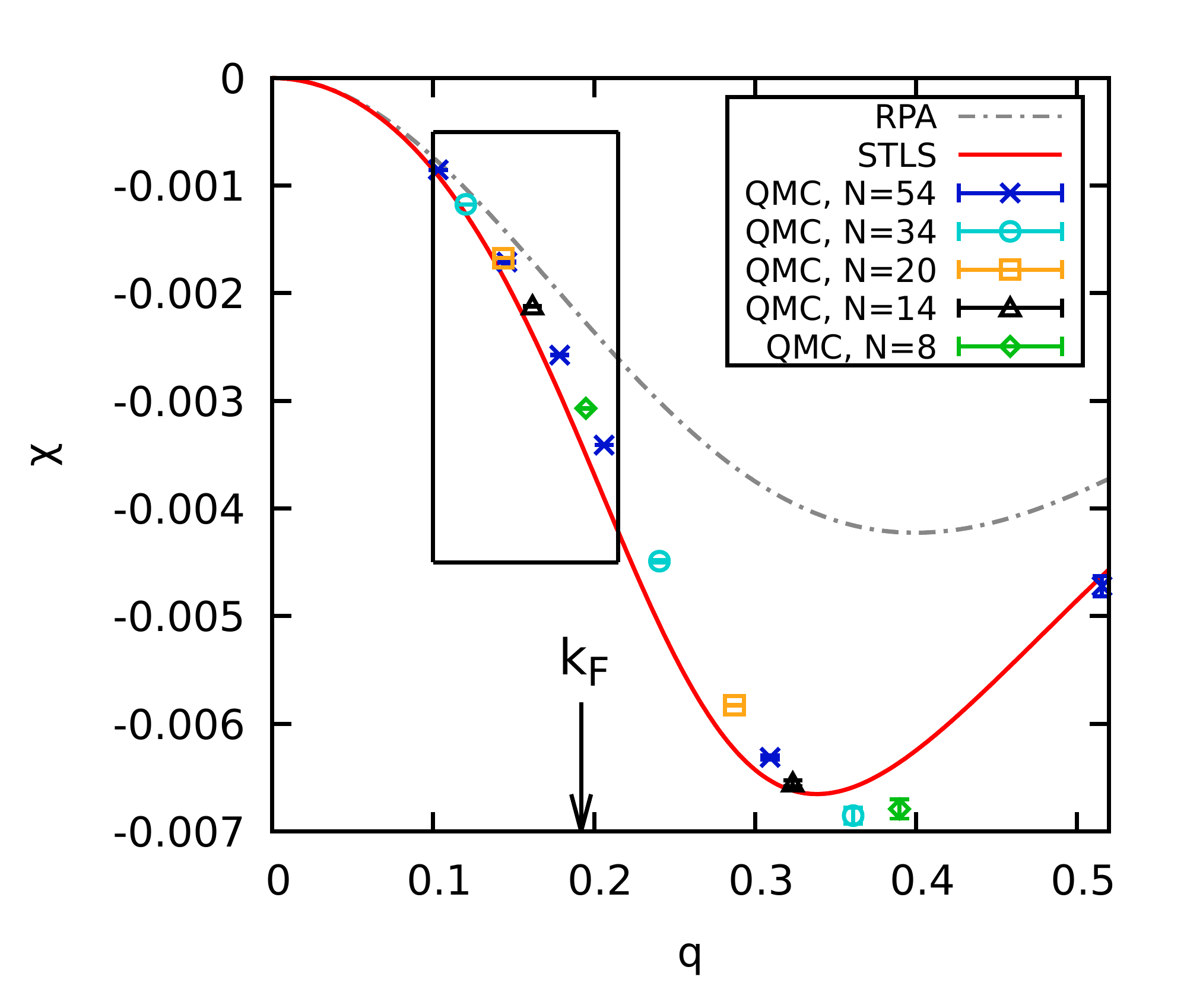}
\includegraphics[width=0.42\textwidth]{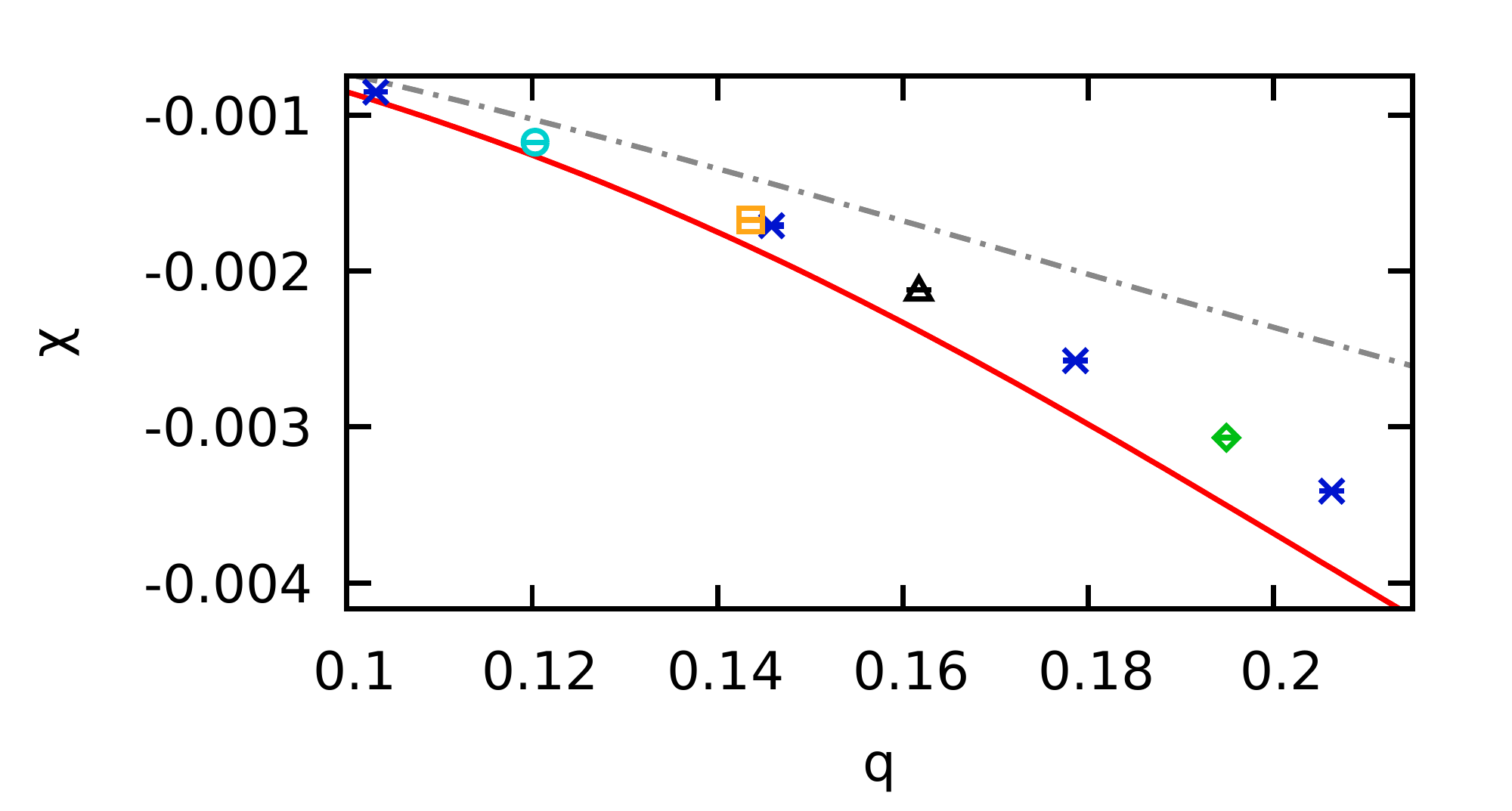}


\caption{\label{fig:kachel_mann}Wave vector dependence of the static response function for the unpolarized UEG at $r_s=10$ and $\theta=1$.
Shown are QMC results according to Eq.~(\ref{eq:weg1}) for different particle numbers (symbols) and the predictions from RPA (grey) and STLS (red). The black arrow indicates the Fermi wave vector, $k_\text{F}=(9\pi/4)^{1/3} / r_s$.
The bottom panel shows a magnified segment.
}
\end{figure}

Due to the momentum quantization in a finite simulation box, QMC calculations are only possible at an $N$-dependent discrete $\mathbf{q}$-grid. Therefore, the investigation of finite-size effects in the static response function requires us to obtain results over a broad wave vector range, as shown in Fig.~\ref{fig:kachel_mann}.
The grey and red curves correspond to the predictions due to the random phase approximation (RPA), cf.~Eq.~(\ref{eq:rpa}), and with a LFC from the (finite-$T$) STLS formalism~\cite{stls,stls2}, respectively. 
For small $\mathbf{q}$, both approximations exhibit the same exact parabolic behavior~\cite{kugler2}. With increasing $\mathbf{q}$, however, there appear significant systematic deviations with a maximum of $\Delta\chi/\chi\sim50\%$ around $q\approx0.35$ (i.e., around twice the Fermi vector $k_\text{F}=(9\pi/4)^{1/3} / r_s$).
The symbols correspond to our QMC results obtained according to Eq.~(\ref{eq:weg1}) and the colors distinguish different particle numbers, in particular $N=54$ (blue crosses), $N=34$ (light blue circles), $N=20$ (yellow squares), $N=14$ (black triangles), and $N=8$ (green diamonds). First and foremost, we note that the main effect of different system size is the $\mathbf{q}$-grid, while the functional form itself is remarkably well converged even for as few as $N=8$ particles, cf.~the bottom panel showing a magnified segment.
This is similar to the analogous behavior of the static structure factor $S(\mathbf{q})$ of the warm dense UEG found in Refs.~\cite{dornheim_prl,dornheim_pop}. Evidently, momentum shell effects as observed at $T=0$ in Refs.~\cite{bowen2,moroni2} do not appear above $\theta=0.5$.
Secondly, we find that the static local field correction due to the STLS closure relation leads to a significant improvement compared to RPA due to the improved treatment of correlations.

We thus conclude that our QMC approach allows, for the first time, to unambiguously assess the accuracy of the multitude of existing and widely used dielectric approximations and, in addition, to provide highly accurate data, which can subsequently be used as input for other theories.
However, a comprehensive study over a broad parameter range is beyond the scope of this work and will be provided in a future publication.

\subsection{Comparison of PB-PIMC to standard PIMC}
\begin{figure}
\includegraphics[width=0.42\textwidth]{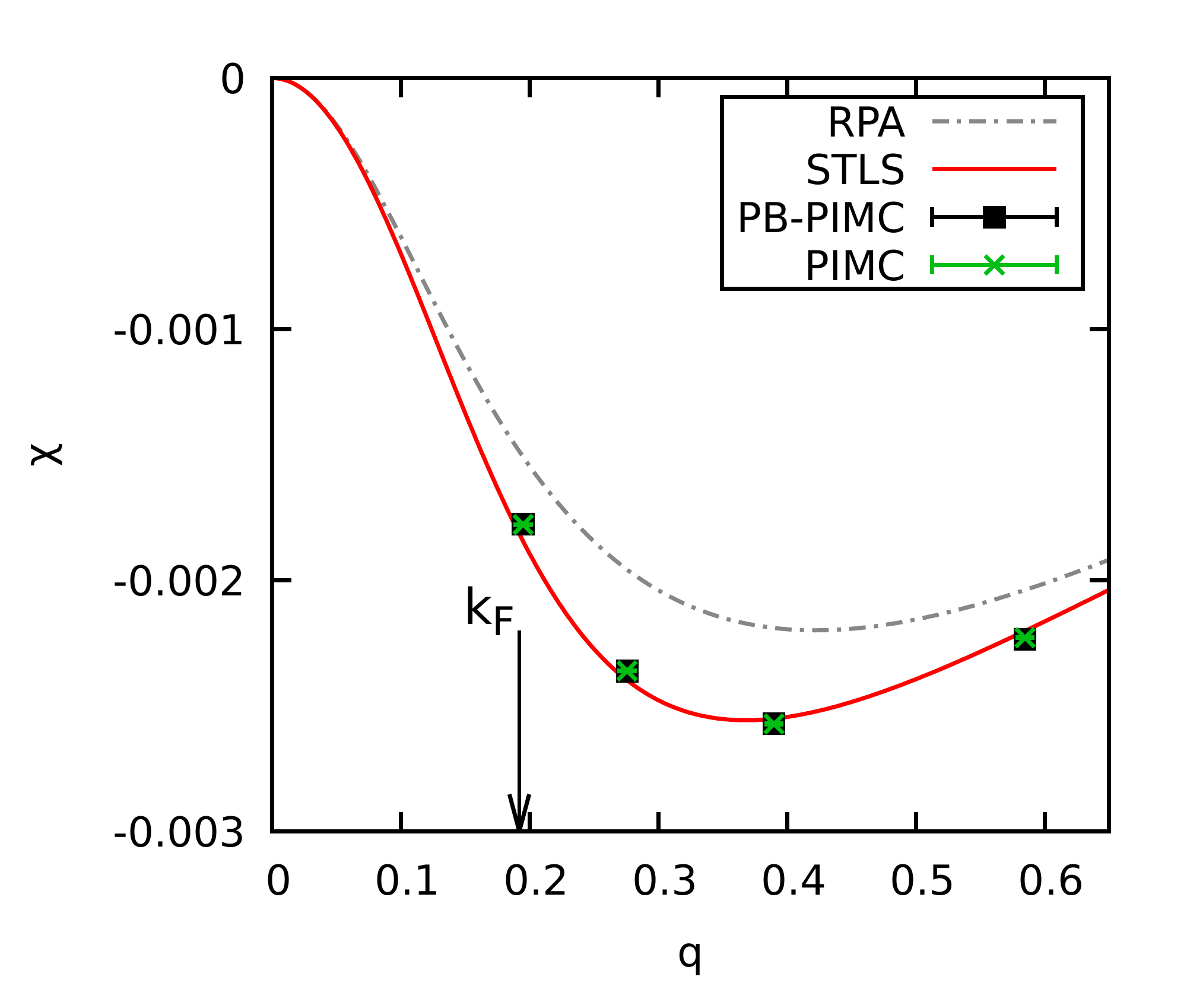}


\caption{\label{fig:std_pb} Wave vector dependence of the static response function for the unpolarized UEG at $r_s=10$ and $\theta=4$.
Shown are QMC results according to Eq.~(\ref{eq:weg1}) for $N=8$ electrons obtained from PB-PIMC with $P=4$ (black squares) and standard PIMC with $P=100$ (green crosses).
As a reference, we also show the predictions from RPA (grey) and STLS (red).
}
\end{figure}

As an additional benchmark for the static response obtained with PB-PIMC, in Fig.~\ref{fig:std_pb} we show $\chi(\mathbf{q})$ for the unpolarized UEG with $N=8$, $r_s=10$, and $\theta=4$. Since for such a temperature fermionic exchange plays only a minor role,  in addition to PB-PIMC (green crosses) also standard PIMC (black squares) calculations are feasible. Evidently, both independent data sets are in excellent agreement over the entire $\mathbf{q}$-range, as expected. 
In addition, we again show results from RPA (grey) and STLS (red) and find qualitatively similar behavior to Fig.~\ref{fig:kachel_mann}. However, due to the four times higher temperature correlations play a less important role, which means that (i) RPA and STLS exhibit less deviations towards each other, and (ii) the density response from STLS is in much better agreement with the QMC data. 
For completeness, we note that a more meaningful assessment of the systemic error due to the STLS approximation requires to eliminate the possibility of finite-size effects in the QMC data (as done in Fig.~\ref{fig:kachel_mann} at lower temperature, $\theta=1$) and, thus, to consider larger particle numbers $N$.

\subsection{Multiple $\mathbf{q}$-vectors from a single simulation\label{sec:mq}}
When we have to perform at least a single (or even a few for different $A$) QMC simulation for each $\mathbf{q}$-value, the investigation of the wave vector dependence as depicted in Fig.~\ref{fig:kachel_mann} is computationally quite involved. 
However, by definition in linear response theory the response of a system to multiple perturbations is described by a superposition of the responses to each perturbation. Therefore, it should be possible to obtain the response function for multiple $\mathbf{q}$-values from a single QMC simulation where we apply a superposition of $N_A$ harmonic perturbations,
\begin{eqnarray}\label{eq:mp}
\hat H_\text{ext} = 2 \sum_{k=1}^{N_A} \left[ A_k \sum_{i=1}^N \text{cos}\left( \mathbf{r}_i\cdot\mathbf{q}_k \right) \right] \quad . 
\end{eqnarray}
The induced density is then calculated for each wave vector $\mathbf{q}_k$ according to Eq.~(\ref{eq:ex}).
Furthermore, the density profile in coordinate space is given by
\begin{eqnarray}\label{eq:multi_n}
\braket{n(\mathbf{r})}_A = n_0 + 2 \sum_{k=1}^{N_A} \left[ A_k \text{cos}\left( \mathbf{r} \cdot \mathbf{q}_k \right) \chi(\mathbf{q}_k) \right] \quad ,
\end{eqnarray}
which means that we have to perform a fit where the free parameters are given by the $N_A$ values of $\chi(\mathbf{q}_k)$.

\begin{figure}
\includegraphics[width=0.42\textwidth]{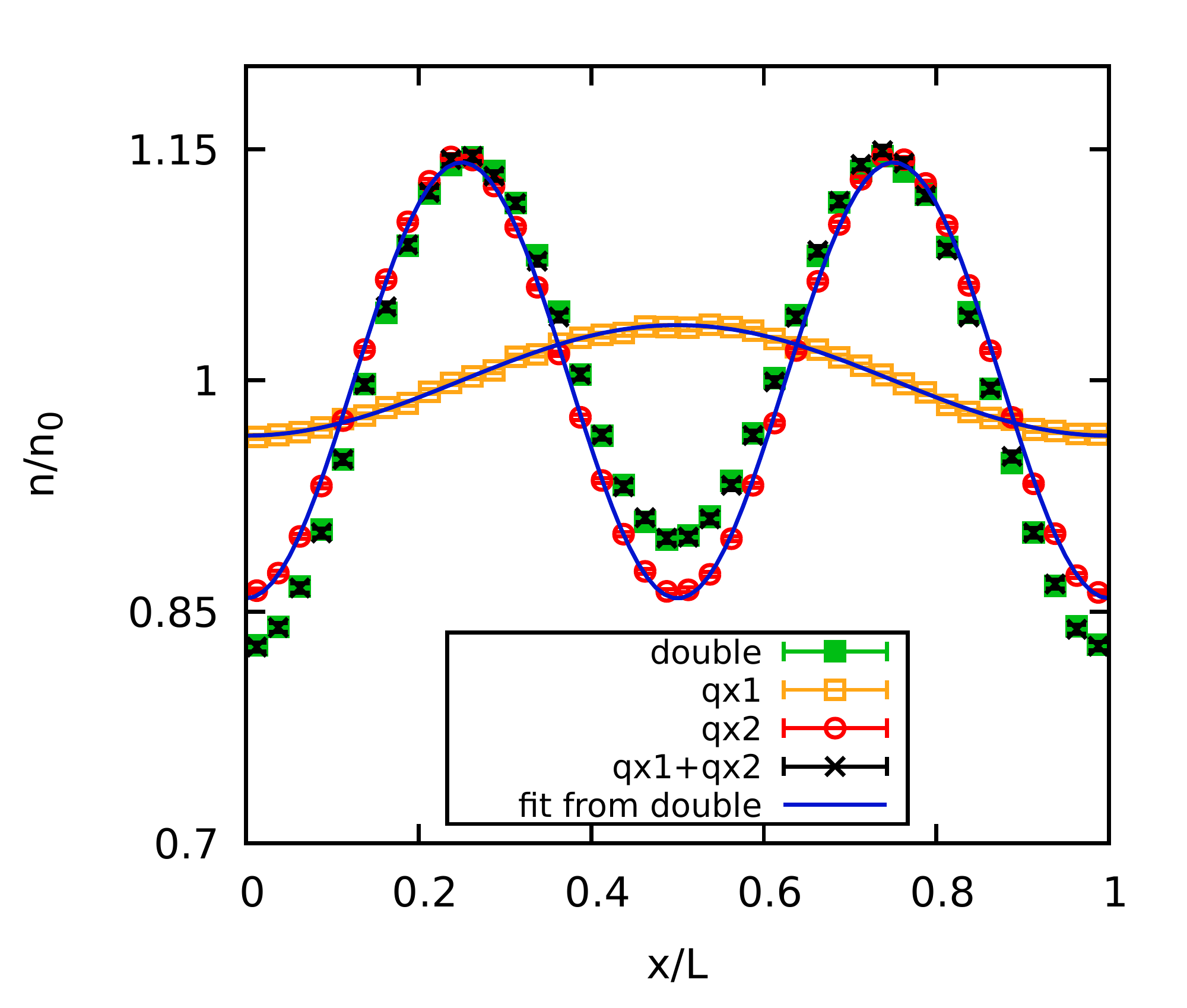}


\caption{\label{fig:double_density} Density profile along $x$-direction for $N=54$, $r_s=10$ and $\theta=1$ with a perturbation amplitude of $A=0.005$.
The green squares correspond to a QMC simulation with a superposition of two $\mathbf{q}$-vectors ($q_x=1$ and $q_x=2$), see Eq.~(\ref{eq:mp}), whereas the yellow and red points have been obtained using two separate QMC simulations each with a single perturbation.
The black crosses correspond to a superposition of the latter two. The blue lines have been reconstructed from a fit to the green squares according to Eq.~(\ref{eq:multi_n}), i.e., by obtaining both $\chi(\mathbf{q}_1)$ and $\chi(\mathbf{q}_2)$ from the density response of the system with two simultaneous perturbations.
}
\end{figure}

In Fig.~\ref{fig:double_density}, we show QMC results for the density profile in $x$-direction for $N=54$, $r_s=10$, and $\theta=1$. The green squares have been obtained from a simulation with a superposition of $N_A=2$ perturbations with $\mathbf{q}_1=2\pi L^{-1}(1,0,0)^T$ and $\mathbf{q}_2=2\pi L^{-1}(2,0,0)^T$ and $A_1=A_2=0.005$, i.e., an amplitude that is expected to be well within the linear response regime.
As a comparison, the yellow and red points correspond to the QMC results with a single perturbation with $q_x=1$ (yellow) and $q_x=2$ (red). Further, the black crosses have been obtained as a superposition of the latter and are in perfect agreement with the green squares. This is a strong indication that the linear response is still valid for multiple perturbations under the present conditions. In addition, we have fitted the RHS.~of Eq.~(\ref{eq:multi_n}) to the green squares and in this way obtained $\chi(\mathbf{q}_k)$ for both $\mathbf{q}_k$-values.
This, in turn, allows us to reconstruct the density response of the system to a perturbation with only a single $\mathbf{q}_k$-value, i.e., the blue curves. Again, we find excellent agreement to the corresponding QMC simulations.

\begin{figure}
\includegraphics[width=0.42\textwidth]{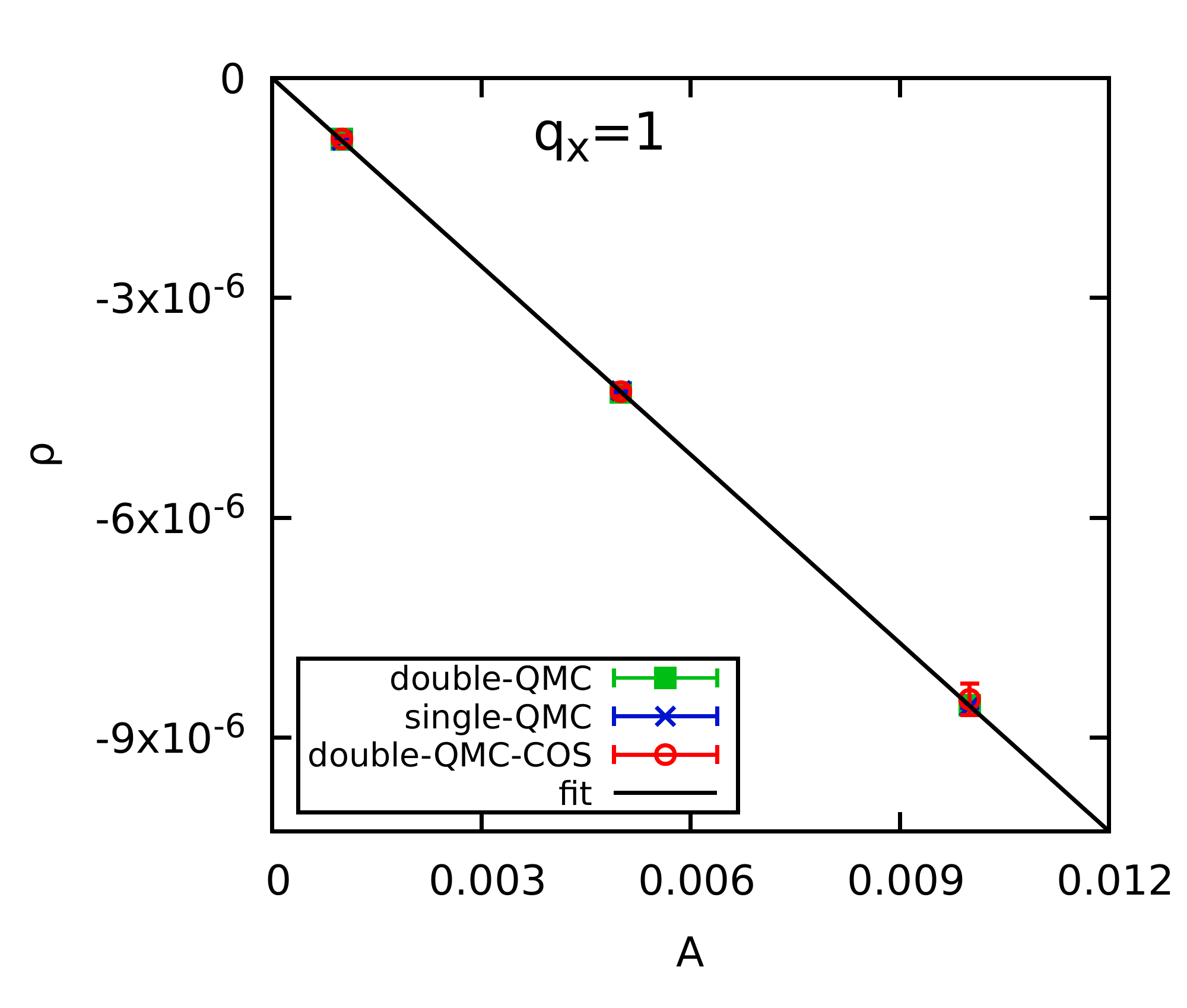}
\includegraphics[width=0.42\textwidth]{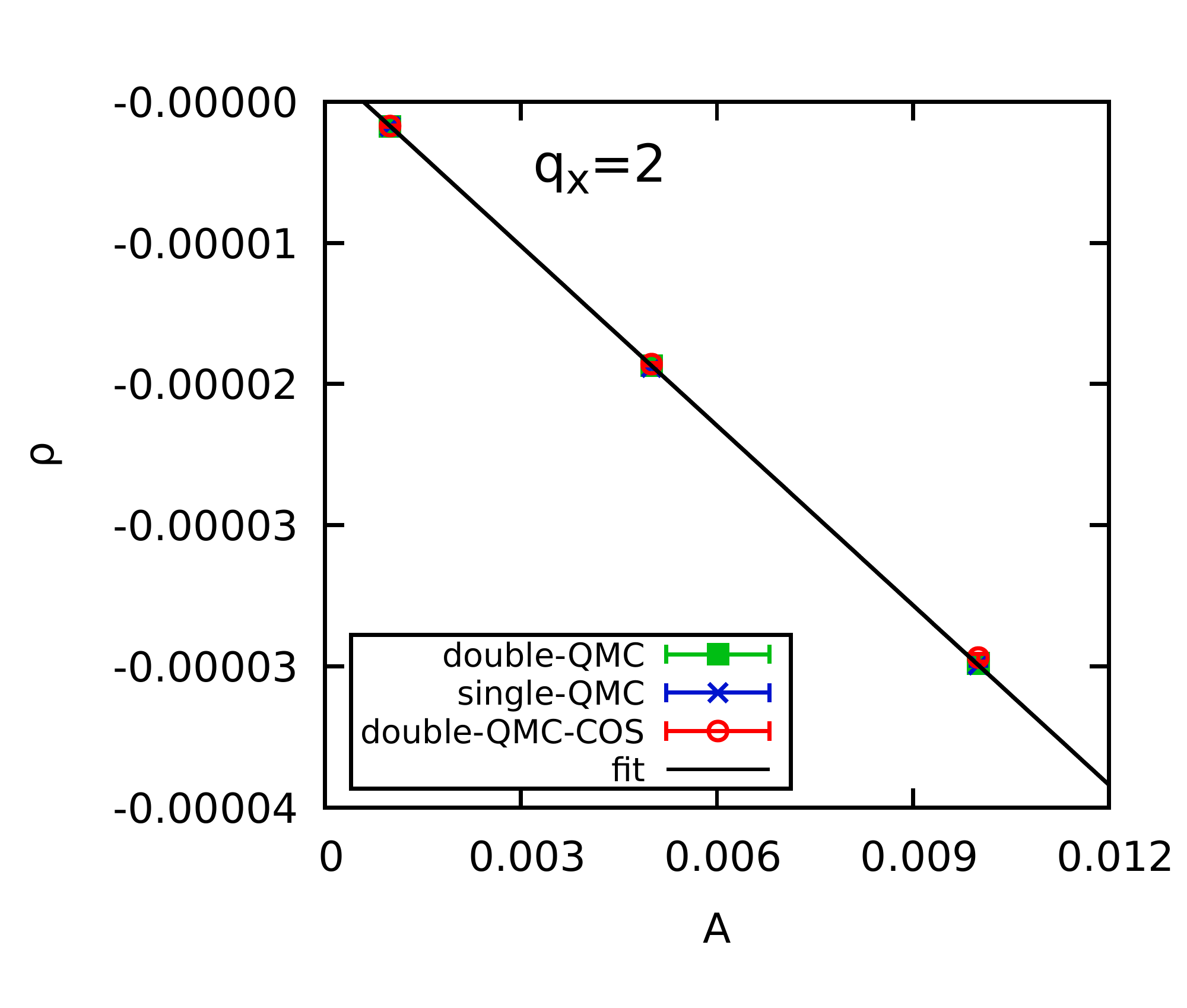}


\caption{\label{fig:double_fitplot}Induced density for $N=54$, $r_s=10$, and $\theta=1$ for a perturbation of wave vector $\mathbf{q}=2\pi L^{-1}(q_x,0,0)^T$.
The blue crosses have been obtained from a QMC simulation with a single perturbation, whereas the green squares and red circles correspond to the direct and cosine-fit results from the simulation with a double perturbation. Finally, the black lines has been obtained by a linear fit to the green squares.
}
\end{figure}

To further pursue this point, in Fig.~\ref{fig:double_fitplot} we show the induced density matrix for different amplitudes $A$. The green squares and red circles have been obtained from a simulation with two $\mathbf{q_k}$-vectors and correspond to the direct QMC estimate and the cosine-fit according to Eq.~(\ref{eq:multi_n}), respectively. The blue crosses have been obtained from the QMC simulation with only a single harmonic perturbation and the red line depicts a linear fit. Evidently, all points are in excellent agreement for all $A$-values both for $q_x=1$ (top panel) and $q_x=2$ (bottom panel).
Therefore, we conclude that it is indeed possible to obtain multiple values of the static density response function $\chi(\mathbf{q})$ simultaneously.

\begin{figure}
\includegraphics[width=0.42\textwidth]{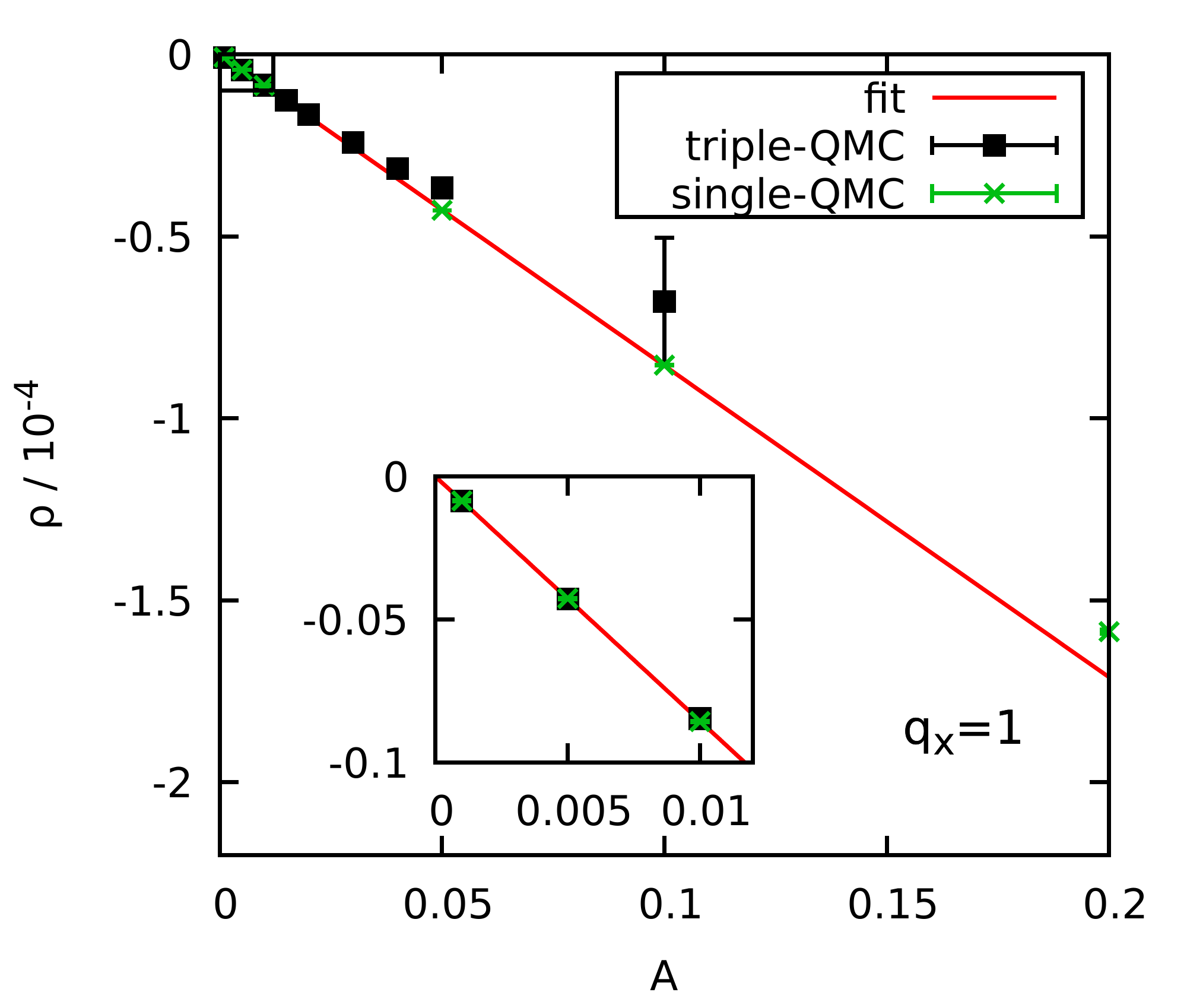}
\includegraphics[width=0.42\textwidth]{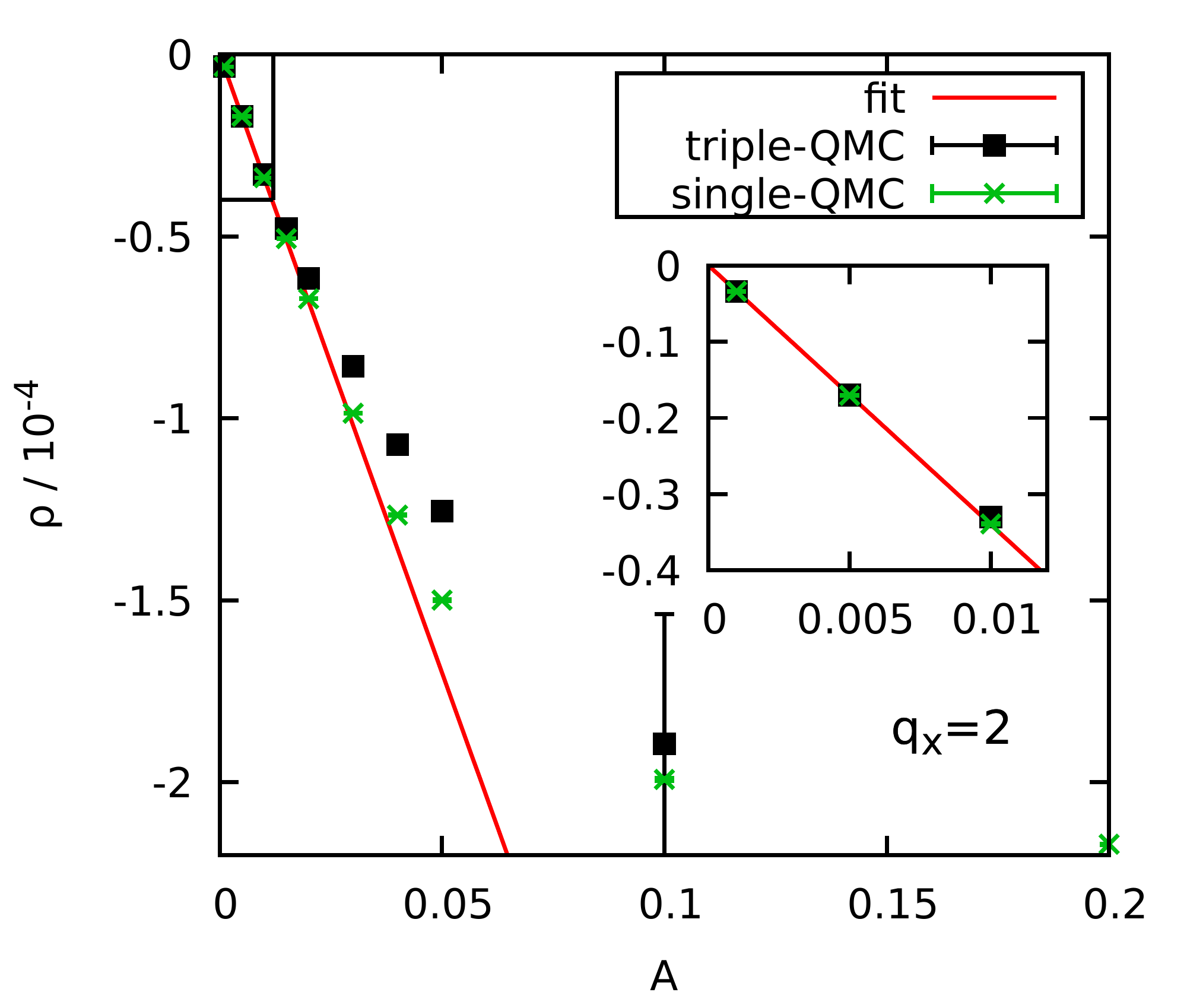}


\caption{\label{fig:triple_fitplot} Perturbation strength dependence for a combination of three wave vectors $\mathbf{q}_i = 2\pi L^{-1} (q_{x,i},0,0)$ with $q_{x,1}=1$, $q_{x,2}=2$, and $q_{x,3}=3$.
The black squares correspond to direct QMC results according to Eq.~(\ref{eq:ex}), the green crosses to direct QMC results from a simulation with a single perturbation, and the red line to a fit in the linear response regime.
}
\end{figure}

Finally, to investigate the perturbation strength dependence for a QMC simulation with a superposition of multiple $\mathbf{q}$-vectors in more detail, we consider a combination of $N_A=3$ perturbations with $\mathbf{q}_1 = 2\pi L^{-1}(1,0,0)^T$, $\mathbf{q}_2 = 2\pi L^{-1}(0,2,0)^T$, and $\mathbf{q}_3 = 2\pi L^{-1}(0,0,3)^T$ and equal amplitude, $A_1=A_2=A_3$, over a broad $A$-range.
The results are shown in Fig.~\ref{fig:triple_fitplot} where direct QMC results for the induced density matrix are shown both from the simulation with the superposition (black squares) and, as a reference, from a simulation with only a single perturbation (green crosses). As usual, the red line corresponds to a linear fit within the linear response regime.
For both $q_x=1$ (top panel) and $q_x=2$ (bottom panel) we observe that the linear response is accurate up to larger $A$. This is expected, since the more perturbations we apply at the same time, the more inhomogeneous the system becomes and, thus, the stronger the total perturbation will be.
Further, we note that this effect is more pronounced for $q_x=2$. This is again a consequence of the larger $\chi(\mathbf{q})$-value which implies that the density response is even larger in this case.

In a nutshell, we find that, while it is possible to obtain multiple $\mathbf{q}$-values of the response function within a single QMC simulation, this comes at the cost that the linear response is valid only up to smaller perturbation amplitudes $A$. However, the smaller $A$ the larger the relative statistical uncertainty of the induced density, which means that there is a tradeoff between more Monte Carlo steps for a simulation with multiple $\mathbf{q}$-vectors or multiple QMC simulations with only a single perturbation and fewer MC steps. In practice, applying a superposition of $N_A\approx3$ perturbations is reasonable.

\section{Summary and Discussion}
In summary, we have carried out extensive permutation blocking PIMC simulations of a harmonically perturbed electron gas to investigate the static density response at warm dense matter conditions. 
To investigate the dependence of the response on the perturbation strength, we varied the amplitude $A$ over three orders of magnitude. For small $A$, linear response theory is accurate and both ways to obtain the response function $\chi(\mathbf{q})$ [i.e., Eqs.~(\ref{eq:weg1}) and (\ref{eq:weg2})] give equal results. With increasing $A$, the system becomes strongly inhomogeneous which leads to a significantly increased sign problem due to the regions with increased density. 
The second important issue investigated in this work is the convergence of the PB-PIMC results for $\chi(\mathbf{q})$ with the number of propagators $P$. For small to medium $\mathbf{q}$, we find that $P=4$ propagators are sufficient at WDM conditions, which agrees with previous findings for the uniform system~\cite{dornheim2,dornheim3}. However, for large $\mathbf{q}$, the external potential exhibits fast spatial variations which lead to increased commutator errors and thus require a larger number of propagators to achieve the same level of accuracy. For the largest considered wave vector, $\mathbf{q}=2\pi L^{-1}(5,0,0)^T$, at $\theta=1$, $r_s=10$, and $N=54$ we find a propagator error of $\Delta\chi / \chi \sim2\%$.
The main effect of system size on the QMC results for the static response function is given by the different $\mathbf{q}$-grid (which is a consequence of momentum quantization in a finite box), whereas the functional form of $\chi(\mathbf{q})$ is remarkably well converged even for small particle numbers. This is in stark contrast to previous findings at zero temperature~\cite{bowen2,moroni2} and can be ascribed to the absence of momentum shell effects at WDM conditions.

Our first brief comparison of the wave vector dependence of $\chi(\mathbf{q})$ computed from QMC to the approximate results from RPA and STLS for $r_s=10$ and $\theta=1$ reveals the stark breakdown of the former when coupling effects are non-negligible. The LFC from the STLS closure relation, on the other hand, constitutes a significant improvement, although there remain significant deviations at intermediate $\mathbf{q}$-values.
Finally, we have investigated the possibility to obtain the static response function at multiple wave vectors from a single QMC simulation. As predicted by the linear response theory, we found that the density response of the electron gas to a superposition of $N_A$ external harmonic perturbations is given by a linear combination of the responses to each of the perturbations. Unfortunately, however, this means that the linear response is valid only up to smaller perturbation amplitudes $A$ as the system becomes increasingly inhomogeneous for multiple $N_A$. Thus, there is a tradeoff between $N_A$ and $A$, and applying a superposition of $N_A=3$ perturbations is a reasonable strategy.

As mentioned in the introduction, accurate QMC results for the static density response function -- and, thus, for the static local field correction -- are of high importance for contemporary warm dense matter research. Based on the findings of this work, the construction of a comprehensive set of QMC results for $\chi(\mathbf{q})$ over the entire relevant $r_s$-range and temperatures $\theta\geq0.5$ appears to be within reach. First and foremost, this will allow one to systematically benchmark previous approximate results for the warm dense UEG, such as STLS~\cite{stls,stls2} (and "dynamic STLS"~\cite{bohm,arora}), VS~\cite{stls2,stolzmann}, or the recent improved LFC by Tanaka~\cite{tanaka_hnc} that is based on the hypernetted chain equation, as well as semi-empirical quantum classical mappings~\cite{pdw_map,pdw_param}. Furthermore, the construction of an accurate parametrization of $G(\mathbf{q};r_s,\theta)$ with respect to $r_s$ and $\theta$ at WDM conditions~\cite{gregori,utsumi,farid} is highly desirable due to its utility for, e.g., new DFT exchange-correlations functionals~\cite{burke2,lu,patrick}, the description of Thomson scattering experiments~\cite{fortmann1,fortmann2}, and the construction of pseudopotentials~\cite{saum1,saum2,pseudo_potential}.
Finally, accurate QMC results for the (weakly and strongly) inhomogeneous electron gas can be used as a highly needed benchmark for different exchange-correlation functionals that are used at WDM conditions~\cite{groth2,gga,gga2,hse,pbe,groth_prl,karasiev2}.

\section*{Acknowledgements}
This work was supported by the Deutsche Forschungsgemeinschaft via project BO1366-10 and via SFB TR-24 project A9 as well as grant shp00015 for CPU time at the Norddeutscher Verbund f\"ur Hoch- und H\"ochstleistungsrechnen (HLRN).


\begin{thebibliography}{26}
\bibitem{fermi_E} In Hartree atomic units, it holds $E_\text{F}=k_\text{F}^2/2$, with $k_\text{F}=(9\pi/4)^{1/3} / r_s$ being the Fermi wave vector.
\end{thebibliography}
\end{document}